\documentclass[12pt,preprint]{aastex}
\shortauthors{Schild, Leiter, \& Robertson}
\shorttitle {Q0957 MECO Model}
\begin{document}
\title {OBSERVATIONS SUPPORTING THE EXISTENCE OF AN
INTRINSIC MAGNETIC MOMENT
INSIDE THE CENTRAL COMPACT OBJECT WITHIN THE QUASAR
Q0957+561}
\author{ Rudolph E. Schild\footnote{Center for Astrophysics, 60 Garden
Street,
Cambridge, MA 02138} , Darryl J. Leiter\footnote{MARC, P.O. Box 7466
Charlottesville, VA 22906} , and Stanley L. Robertson\footnote{
Dept. of Physics, Southwestern Oklahoma State University,
Weatherford, OK 73096}}
\begin{abstract}
Recent brightness fluctuation and auto-correlation analysis of time
series data, and micro-lensing size scales, seen in Q0957+561 A,B,
have produced important information about the existence and
characteristic physical dimensions of a new non-standard
magnetically dominated internal structure contained within this
quasar. This new internal quasar structure, which we shall call the
Schild-Vakulik Structure, can be consistently explained in terms of
a new class of gravitationally collapsing solutions to the Einstein
field equations which describe highly red shifted, Eddington
limited, Magnetospheric, Eternally Collapsing Objects (MECO) that
contain intrinsic magnetic moments. Since observations of the
Schild-Vakulik structure within Q0957+561 imply that this quasar
contains an observable intrinsic magnetic moment, this represents
strong evidence that the quasar does not have an event horizon.
\end{abstract}
\section{Introduction}
Analysis of gravitational micro-lensing observations of the quasar
Q0957+561 (Schild and Vakulik, 2003; Schild, 2005) has offered
strong evidence for the existence of an intrinsic structure within
this quasar that can be only be explained by a non-standard
luminous quasar model consisting of a thin accretion disc whose
interior is essentially empty of matter out to a luminous inner edge
which resides at about 70 gravitational radii from the central
compact object, and which is also surrounded by an order of
magnitude larger outer ring-shaped Elvis Structure where the broad
blue-shifted emission lines are formed. While this non-standard form
of inner quasar structure (which we shall call the Schild-Vakulik
Structure) was shown to have the remarkable property of being able
to explain all of the observed features of Q0957 obtained from 24
years of gravitational micro-lensing, the physical origin and
interpretation of this exotic inner quasar structure remained an
unsolved mystery. During the same time period that the gravitational
micro-lensing based discovery of the Schild-Vakulik structure in
quasar Q0957 occurred, the related discovery of a new paradigm in
astrophysics based on X-ray and Radio Astronomy data had begun to
emerge. This new paradigm came in the form of five sequentially
published papers which presented strong observational and
theoretical evidence that both galactic black hole candidates (GBHC)
and Active Galactic Nuclei AGN have observable intrinsic magnetic
moments.

This latter discovery was revealed in the following manner: a) First
it was argued (Robertson and Leiter, 2002) that the spectral state
switch and quiescent luminosities of low mass x-ray binaries, (LMXB)
including GBHC, can be well explained by a magnetic propeller effect
that requires an intrinsically magnetized central object. b) Second
it was shown (Leiter and Robertson, 2003; Robertson and Leiter,
2003) that this result was consistent with the existence of a new
class of gravitationally collapsing solutions of the Einstein field
equations in General Relativity which describe highly red shifted,
Magnetospheric, Eternally Collapsing Objects (MECO) that do not have
trapped surfaces leading to event horizons. These general
relativistic MECO solutions were shown to emerge from the physical
requirement that the structure and radiation transfer properties of
the energy-momentum tensor on the right hand side of the Einstein
field equations for a collapsing object must contain equipartition
magnetic fields that generate a highly redshifted Eddington limited
secular collapse process which satisfies the Strong Principle of
Equivalence (SPE) requirement of time like world line completeness.
 
Third it was demonstrated (Robertson and Leiter, 2004) that GBHC
and AGN modeled as intrinsically magnetic MECO objects can produce
jets that emit radio-infrared luminosity correlated with mass and
x-ray luminosity in a manner which correctly predicts the observed
exponent, mass scale invariant cutoff, and radio luminosity ratios
of both GBHC and AGN. Finally a full discussion of the entire
observational and theoretical program for the "Magnetospheric
Eternally Collapsing Object (MECO) model of Galactic Black Hole
Candidates and Active Galactic Nuclei" was published (Robertson and
Leiter 2005) as chapter one in the book "New Directions in Black
Hole Research" recently published by Nova Science Publishers. In the
following sections we show that the two different lines of research
discussed above converge in a manner allowing the construction of a
logical chain of observational and theoretical arguments which
solves the mystery of the physical origin of the Schild-Vakulik
structure observed within the quasar Q0957. In order to do this we
use the Magnetospherical Eternally Collapsing Object (MECO) paradigm
(discussed in detail in the Appendices of this paper) as a model to
interpret the physical meaning of the gravitational micro-lensed
Schild-Vakulik structure seen in Q0957. In this context we show that the
Schild-Vakulik structure actually represents the key observable
dynamic signature of a 3.6 billion solar mass MECO acting as the
central compact object in this quasar instead of a black hole. Hence
we are led to conclude that the observation of the Schild- Vakulik
structure within Q0957 represents strong evidence that this quasar
does not have an event horizon.

\section{Quasar Structure Size Estimates from Brightness Time Series
Auto-correlation}

The most accurate direct measurement available of the inner quasar
size is in the long time series brightness data collected over many
years to study the gravitational lens time delay and micro- lensing.
Evidence for the outer ring (Elvis) structures in this data set has
already been discussed by Schild (2005) from analysis of the
internally created repetitions seen in the long brightness record.
In addition an inner structure has also been discussed in Schild
(2005) and in previous references. Heretofore it has been discussed
as the ``luminous inner edge of the accretion disc'' according to
standard ideas of quasar structure dominated by a black hole
surrounded by an accretion disc and perhaps outer clouds or Elvis
outflow structures. However recent more accurate numerical estimates
of the size of the inner quasar structure of Q0957 have implied that
its size is actually larger than previously estimated and for this
reason better explained as the luminous ring expected at the
magnetospheric radius of an empirical quasar model in which the
central compact object in the quasar is dominated by the dynamic
effects of an intrinsic magnetic propeller acting in this sensitive
inner region.This improved more accurate empirical quasar model 
for Q0957+561 has
been found to be most consistently explained in the context of the
recently discovered new class of gravitationally collapsing
solutions to the Einstein field equations which describe compact
gravitationally collapsing object in terms of highly red shifted,
Magnetospheric, Eternally Collapsing Objects (MECO) that do not have
trapped surfaces leading to event horizons. In the following
paragraphs, in order to help the reader of this paper maintain the
continuity and flow of the observational arguments which support the
above conclusions, the specific details about the MECO model and the
published articles associated with it have been put into a series of
short appendices Appendix 1-11 and will be referred to when needed in
order to clarify the details of specific
calculations associated with the observational data.

We now begin discussing the observational arguments referred to
above by first noting that two manifestations of the dynamic inner
structure of Q0957+561 A, B have been seen: 1. Direct
auto-correlation calculations of the brightness record for the two
image components have revealed an inner structure with an associated
time scale of approximately 10 days (Schild 2005). 2. Time delay
calculations by several research groups have indicated that several
time delays seem to be present. In particular, in addition to the
now firmly established 417.1 day time delay (Colley et al 2003)
there is convincing evidence for lags of 404 days (Schild, 1991;
Thomson and Schild 1997) and 424 days (Pijpers 1997, Oscoz et al
2002, Pelt et al 1996)). It is likely that these lags caused by
internal structure are part of the reason why it has been so
difficult to determine the cosmological time delay. Regarding 2
above, recall that the time delay controversy in the 1990's era was
probably a manifestation of quasar structure causing several lags to
produce auto-correlation peaks that can also create spurious
cross-correlation peaks.
 
Recall that the difference between the 404-day concurrent time delay
value (Schild 1991; Thomson and Schild 1997) and the erroneous Press
et al (1992) value of 534 days is exactly the 129 day value found as
an auto-correlation peak illustrated in Fig. 1 of Schild 2005.

The existence of internal reflections/fluorescence
as manifestations of internal
quasar structure can easily be seen in the most recent brightness data for
a different quasar, Q2237, plotted at:\\

\textbf{http://www.astrouw.edu.pl/~ogle/ogle3/huchra.html.}\\
Here we can recognize two distinct kinds of patterns in the brightness
fluctuations of the four quasar images. Because the time delay between
arrivals of the four images is essentially 0 days (see Vakulik et al, 2005,
astro-ph/0509545), the pattern of nearly identical fluctuations are
clear manifestations of
intrinsic quasar brightness fluctuations. A non-intrinsic
(microlensing) brightness fluctuation, peaking at JD-2450000 = 3500
is seen for image B (yellow points).
We now focus on the pattern of intrinsic brightness fluctuations.
Local maxima are recognized at JD -2450000 = 2950, 3300, and 3700, with 
a possible
first peak around 2500 poorly observed because of the poor OGLE III
data sampling. The amplitudes and durations of these peaks, around
0.4 magnitudes for 150 days, are comparable to the peaks and lags
identified in Q0957 (Schild, 2005 and earlier references). (Recall that the
redshift of Q2237 is 1.69, about 20 percent higher than that of Q0957,
so time dilation correction improves the comparison.)

Thus the two gravitational lens systems with adequate data show a complex
behavior whereby an intrinsic fluctuation is seen at multiple times,
almost certainly betraying the existence of complex internal quasar
structure that can confound time delay determination but that also
provides important clues about the structure of the quasar.
The previous generation of quasar models has ignored these structures and
modeled the lensed quasars as solid accretion discs. This has produced
in all the
numerous available simulations of Q2237 microlensing,
the contradictions that an accretion disc small enough to allow
interpretation of microlensing events as due to solar mass stars produces
brightness fluctuations with 2 to 3 magnitude amplitudes, and transverse
velocity estimates 10 times larger than estimates from normal cosmology.
These models have also produced the failed prediction of Wyithe, Webster,
and Turner (2000) that now drives us away from the simple accretion
disc model.

The Q0957 structures discussed above
have durations and lags of order 150 days
(observer's clock) or 60 proper days and are presumed to originate in the
continuum emitting regions associated with the Elvis structures (Schild
2005). Shorter time scale lags, also caused by quasar internal structure,
allow us to estimate the size of the inner edge of the accretion
disc, using methodology and results from Schild 2005. We assume that
the cosmological time delay is 417.1 days (Colley et al 2003) but
that the internal quasar structure also produces lags at t1 and t2
following the initial impulse.

Thus the impulse seen in the first arriving A image will correlate
strongly with the second arriving B image at lags of 417, 417+t1,
and 417+t2 days. Of course there should be strong correlation at
417-t2 and 417-t1 days due to secondary impulses contained in the A
(first arriving) image. Evidence for these lags shorter than the
cosmological time delay has been reported in Schild and Thomson
(1997) who note that "this calculation produces a broad maximum with
an absolute peak for 387-day lag. Note that the calculation produces
a number of peaks near 400 days, and the peaks have a uniform
spacing of 16 days, which may correspond to an internal reflection
within the inner quasar structure ..." Because the hot inner radius
of the accretion disk will be seen as the small brightness
enhancement for 2 lags corresponding to the ring radius and for the
known inclination, we ask if the measured lags for the 404 and 424
day measured delays follow from the Schild (2005) model and geometry
in Q0957. The answer is an unqualified yes. The shortest lag, 7 days
observed (ie, 424 - 417 days) must correspond to the interaction
(reflection/fluorescence) from the near side
of the ring. Then the reflection
from the far side must be at 13 days, which is both the lag measured
(ie, 417 - 404 days) and the lag predicted given the preferred case
1 geometry of Schild 2005.

In other words, given the quasar inclination measured
by Schild (2005) from the Elvis
structures, and given one of the measured inner ring lags, we can
predict the second (measured, 13-day)inner ring lag. This quasar
structure and its interpretation is necessarily complicated because
the effects of the solar mass micro-lensing star must selectively
magnify the 2 sides of the inner edge of the accretion disk
differently at different times. While this produces complexity in
interpretation, it also generates information about the direction of
motion of the micro-lensing star in the lens galaxy G1. Finally,
from this complex chain of analysis of internal quasar structure, we
determine the size of the luminous inner edge of the accretion disk.
For the geometrical factors of the model as determined above, the
measured 424-day and 404-day lags combined with a 417.1-day
cosmological time delay (Colley et al 2003) and correction for the
54 degree inclination of the quasar's pole from the line of sight
(Schild 2005) gives a rather large value for the inner radius of
the accretion disk of $(3.9 \pm 0.16)\times10^{16}$ cm, where the error
is estimated from the 1-day discreteness of the reported lags and the
averaging of the two estimates from the above two lag determinations.

\section{Empirical Estimates of the Thickness of The Hot Annular Ring
Observed To Occur At The Inner Radius of The Accretion Disk}

The inner radius of the accretion disk (which will later be
associated with the MECO magnetospheric radius) is considered to be
the smallest observable structure in the Q0957 quasar and therefore
the rapid fluctuations observed by Colley and Schild (2003) are
presumed to originate there. However it is not yet clear by what
mechanism the brightness fluctuations are produced.
We consider that there are two probably two processes contributing to the
observed brightness patterns, micro-lensing and changes in the quasar
structure. The most critical observations define a small size
limit as seen in the CS03 data, which seem to demonstrate rapid
brightness fluctuations of 1 percent amplitude and 12 hour duration
in both the intrinsic quasar brightness fluctuations and in the
microlensing.

We refer to the patterns of fluctuations shown in Fig. 1 of CS03.
There, on Julian Dates 2449701 - 05 we see a pattern of fluctuations
which was seen the first year of observations in image A (filled
dots) and after a 417-day time delay, seen in image B (open
circles). The brightness trend seems to show convincingly that a
pattern of intrinsic fluctuations was seen, with several brightness
changes of approximately 1 percent with a time scale on the order of
12 hours. Thus in day 1, the quasar appears to have declined in
brightness by .017 magnitudes in 0.43 days (solid line fitted to the
data points in the upper panel). Note that this must be an intrinsic
quasar brightness fluctuation, because it is seen in both the first-
and second-arriving images. In addition, micro-lensing fluctuations
of comparable amplitude must exist, as again illustrated in Fig. 1
of CS03. Referring to the data for Julian Day 2449706, we can see
from the many time- overlapping points that the quasar image
brightness does not agree for this date, and the difference between
the two observations for the same proper quasar time shows the
profile illustrated in the lower panel of Fig. 1, where a
micro-lensing event with 12 hour time scale and .01 magnitude
amplitude is detected. Because this profile must originate in
micro-lensing, it would be expected to have a cusp-shaped profile
unless the source structure is partially resolved, as seems to be
observed here.

These simple amplitude and time-scale estimates of micro-lensing and
intrinsic quasar brightness changes are entirely compatible with
previously published estimates, expressed as wavelet amplitude
(Schild 1999), structure function (Colley and Schild 1999), and
Fourier power spectrum (Thomson and Schild 1997).
The wavelet fits to the A and B images separately
(Schild 1999) measure the total amplitude in fluctuations
independently of whether the fluctuations originate in intrinsic or
micro-lensing variability. Nevertheless the amplitudes measured for
the A and B images are seen in Schild 1999 Fig. 8 to be of amplitude
$1\%$ on time scales of 2 days in the mean. The event of CS03 is
exceptional by a factor of 4 in time, but this is of course a
probable selection effect. The view that these most rapid low
amplitude fluctuations originate in the inner ring of the accretion
disk produces a new view of the nature and origin of the quasar
luminous signal and its micro-lensing.

If we adopt for the moment a model of the inner quasar where a
disturbance of some kind, perhaps related to ingestion of a mass
unit, and seen throughout the inner corona and its surrounding inner
ring of the accretion disk, then we would expect to see the
brightness enhancement signaling this event originate in the corona
and again, after short lags of order of days, when the event is seen
in the nearer and then the farther sides of the inner ring of the
accretion disk. Because each of these regions will be microlensed
independently and differently, it would be unsurprising if the
micro-lensing had the same amplitudes and time scales for brightness
fluctuations as the intrinsic brightness fluctuations, as observed.
This important possible kind of micro-lensing is significantly
different from the simpler view that the fluctuations are just
caused by luminous elements crossing the caustic pattern originating
in the lens galaxy (Gould and Miralda-Escude, 1997; Rauch and
Blandford 1991). However our new approach requires the quasar to
have very fine structure, whose size we now estimate.

\subsection{ Estimation of Source and Micro-Lensing Amplitudes and
Their Time Scales}

We estimate first the dimensions implied if we adopt the simple
argument that for a significant brightness fluctuation to occur,
something must be altering its physical properties and therefore its
luminosity on a time sufficiently long that the the observed
coherence does not violate the principle of causality (meaning that
the event is limited in size by the velocity of light). Thus the
proper scale of of the quasar emitting region that coherently
changed its luminosity 5\% (because dilution by the steady luminous
emission of the large outer Elvis structures produces 80\% of the
observed quasar brightness according to SV03) is at least as large
as the light travel distance, $ct$. For the CS03 event with observed
time 12 hours, and proper time 12/2.41 hours = 5 hours, the probable
diameter of the coherent light emitting region is $5.4\times10^{14}$
cm. The numerical value of this length, estimated from the coherent
brightness fluctuations, represents the radial thickness of the hot
annular band at the inner radius of the accretion disk.

\subsection{ Micro-lensing By Luminous Matter Passing Behind 
Microlensing Cusps}

\textbf{ 3.2.1 In-fall case}

A second estimate of fluctuation source size comes from the model
that the fluctuation is due to some luminous unit crossing a cusp
originating in the lens galaxy. As already noted by Schild (1996),
the cusp pattern would be produced by planetary mass micro-lensing
objects, which amplify the shear introduced by the stars also known
to be in the lens galaxy, but inferred not to be sufficient to be
the baryonic dark matter. This process has already been modeled by
SV03 from which we can adopt some principal results. In SV03 the
shear due to solar mass stars is modeled as originating in a
population with 0.1 solar mass, but this adopted value is not
expected to appreciably affect results since the planetary mass
population produces a very significantly finer cusp pattern with a
factor 10 larger optical depth. The model adopts standard values for
cosmology and for the transverse velocity. A principal result of the
model is that the micro-lensing brightness fluctuations that would
be observed should be typically .01 magnitudes on a time scale of 10
days (observer's clock), for an adopted thickness of the inner ring
of $5.4\times10^{14}$ cm and for $10^{-5}M_\odot$ micro-lensing
particles. This fluctuation time varies as the size of the emitting
region and as the square of the micro-lens mass. Thus the structure
size scale estimated would be $10^{13}$ cm and the mass would be
$10^{-7} M_\odot$ for the somewhat resolved micro-lensing event
found in CS03.

We find it unlikely that the Q0957 quasar has coherent structure on
such small size scales, because the brightness fluctuations have no
hint of it, although it is possible that an ingested mass unit
powering the quasar does. Based on the above discussion, we adopt
for now the more conservative conclusion that the observed rapid
brightness fluctuations result from a local brightness change in the
accretion disk seen at multiple epochs.

\textbf{3.2.2 Micro-lensing By Orbiting Luminous Material Generated
by Magnetospheric Effects}

Another micro-lensing scenario has been suggested by Gould and
Miralde-Escude (1997) and previously by Rauch and Blandford (1991).
In this scenario, orbiting blobs created in the magnetospheric ring
region of the inner accretion disc,[e.g. like that described by the
Magnetically Arrested Disk (MAD) model (Igumenschev, Narayan, and
Abramowicz, 2003) would pass behind the cusp pattern originating in the
lens galaxy. However, to be visible, such blobs would necessarily
have to have a luminosity comparable to that of the quasar coming
from a volume with the diameter of the sun. Furthermore, these
luminous blobs would tend to make asymmetrical brightness profiles,
characterized by successive brightness peaks only, contrary to the
wavelet result by Schild (1999) that equal positive and negative
fluctuations are found. Furthermore, the mechanism would produce
highly periodic brightness effects not observed in any of the
lensed quasar systems. Hence we consider this mechanism to be not
an important indicator of the quasar structure in Q0957 (for more
details on this issue see the theoretical analyis discussion in
section 7 below).

\section{Structure Function Estimates Inferred From Long Term Brightness
Fluctuations}

With the shortest time scale brightness fluctuations understood to
be indicative of the thickness of the hot inner annulus of the
accretion disk in Q0957+561, we consider next the properties of
brightness fluctuations observed on longer time scales. The increased
amplitudes of brightness fluctuations on longer time scales is often
described as the structure function, with amplitude of fluctuation
expressed by some measure such as rms deviation, as a function of the
interval of time between successive samples of this fluctuating quantity.

Estimates of the structure function have been given for time scales
of 1 day to years in Colley and Schild (1998), or in a wavelet calculation
and representation by Schild (1999), and also as a Fourier
representation of the micro-lensing component (Schild 1996, Fig. 4;
Thomson and Schild 1997.) This last reference considers only
fluctuations on long time scales, approximately 100 days, and will
not here be considered further. In the direct structure function
estimates of CS03 and Colley and Schild, 1998,
the variance quantity plotted is the
square of the brightness difference amplitude as a function of lag
between brightness samples. Thus for the plotted fit in Fig. 5 of
Colley and Schild (2000),
the mean fluctuation amplitude for 1-day lag is 0.0063 mag,
and the square root of the variance is proportional to the lag. In
other words, the mean brightness increase expressed as a rms, is a
linear function of lag time.

The same information has been gleaned from a wavelet analysis of the
A and B image brightness records as described in Schild 1999. There
we report in Figs. 5 and 6 the wavelet amplitude expressed as a mean
absolute deviation, and in Figs. 7 and 8 the amplitude expressed as
an rms deviation, which is comparable to the presentation in
Colley and Schild (2000).
The results agree well, with the rms deviation extrapolated to 1 day
lag of 0.005 mag, and a linear trend of rms deviation increasing
with lag for lags up to approximately 20 days. This linear increase
in the structure function is significant and has never been
interpreted in the context of a physical process, probably because
it has heretofore been anomalous and unexpected. Standard accretion
disc models would adopt a simple picture that a disturbance in the
accretion disc limited by the causality principle, with the amount
of the brightness change increasing with time as the disturbance
spreads, would produce a structure function that increases with the
square of the time scale for an optically thick accretion disc where
the observed optical disturbance increases as the area within
causality, or as the cube for the optically thin case.

The observed linear increase is, however, compatible with the MECO
quasar model discussed in section 7, where the emitting source is
seen as an "annular band" around the equator of the compact object,
with the band "thickness" or cross-section diameter of
$5.4\times10^{14}$ cm, as estimated above. As the disturbance
propagates along the length of the band, the causally connected luminosity
increases linearly. However, note that the linear increase should
end when the entire length is in causal connection; this occurs when
light has traveled a distance of 30 proper light days, the ring
diameter, corresponding to an elapsed time of 14 light days (in the
observer's frame) when allowance is made for cosmological effects
(the [1+z] correction to time) and for geometrical effects related
to the orientation of the quasar. Thus the "events" seen in wavelet
representation would produce up to 16-day wavelets. This 16-day
limit may be present in the measured wavelet amplitudes, Figs
5,6,7,and 8 of Schild 1999, where we see that the linear trend for
2,4,and 8-day wavelets curves toward a shallower slope in passing to
16, 32, and 64-day wavelets.

The upward trend of amplitude for image B wavelets at 64 days
probably results from the micro-lensing events predicted by the SV03
model to relate to the luminous quasar Elvis outer structures. Thus
we conclude that both the basic time scale found for rapid quasar
variability and the linear increase according to the measured
structure function seem to favor an origin in an inner band-like
MECO ring with approximately the theoretically expected thickness
and radius.

\section{Time Scales for Quasar Fueling Events Associated With
Unit Mass Ingestion}

Fundamental to the understanding of the Quasar luminosity as a
response to fueling must be an analysis of time scales associated
with the fueling process. We begin with some comments about the
nature of mass condensations in the universe.

A fundamental precept of the modern astrophysical view is that the
universe is dominated by a non-baryonic "Cold Dark Matter" (CDM)
which seeded and developed all structure observed today. We consider
the arguments for the existence of this component weak, because the
matter has not been found in laboratories despite 15 years of very
determined searches, and because observations of the low-redshift
(local) universe do not find the expected substructure (Putman and Moore
2002) and because observations at highest redshifts at which
galaxies and quasars are observed do not conform to predictions. In
particular, at redshifts 4-5, fully assembled galaxies with $10^{9}
M_\odot$ are found far in excess of simulations, and clustering of
quasars and galaxies at redshifts 3 - 5 is much stronger than
simulated in all CDM simulations. Worse, the high redshift galaxies
and quasars are observed to have solar metal abundances, even though
the epoch of maximum star formation is expected to be between
redshift 1 and 2. And contrary to the CDM theory, dwarf galaxies do not
have the cusp-like central structure expected (Spekkens, Giovanelli, and
Haynes, 2005).

The available simulations and theory consistently
conclude that on spatial and mass scales relevant for the quasar
fueling, structure in the CDM distribution would be unimportant. A
possible exception is the report by Diemand, Moore, \& Stadel (2005)
that CDM clustering on planetary mass scales is observed in
simulations for axion Cold Dark Matter.

For the remainder of this report, we adopt the point of view that
cold dark matter probably is too diffusive to contribute
significantly to the fueling of quasars by discrete mass units. On
the other hand, the baryonic dark matter is now known to be
significantly aggregated as condensations on specific scales. Stars
seem to be in evidence everywhere, and their mass is probably
reasonably taken to be in a log-normal distribution with a most
probable value of $0.5 M_\odot$ and a half-with of a factor 10
around this most probable value (note that for a log-normal
distribution, the mean, mode, and median are not the same, unlike
the commonly accepted result for Gaussian statistics). The baryonic
dark matter has been identified from quasar micro-lensing as a vast
network of "rogue planets" having planetary mass and populating
interstellar space everywhere (Schild, 1996, 2004b).

A hydrodynamical theory that predicts this population (Gibson 1996,
1999) as ``Primordial Fog Particles'' ascribes their origin to
fossil fluctuations pervading the universe and forming planetary
mass condensations at the time of recombination, 380,000 years after
the Big Bang. Since their discovery and interpretation from quasar
micro-lensing, they are now being seen in quasar "extreme scattering
events" (Walker and Wardle, 1998; Wardle and Walker, 1999)
and in "Pulsar Scintillation
Scattering" events (Hill et al, 2004). The same hydrodynamic theory
that predicts the formation of the "Primordial Fog Particles"
(Gibson 1996) also predicts that nature aggregated matter at the
time of recombination in globular cluster mass scales, $10^{6}
M\odot$. These would be the objects mysteriously appearing on short
time scales by the thousands during galaxy-galaxy collisions. Nearly
all theories of structure formation acknowledge this ``Jeans mass''
scale of expected structure formation at recombination. This leaves
us with a picture of baryonic dark matter having only a small
fraction in smoothly distributed gas form, and more importantly
aggregated on scales of planets, stars, and globular clusters.
Thus we accept these as the unit fueling components for qusaars.

The quasar response to this fueling process, seen as brightness
fluctuations, has been summarized by De Vries et al (2005),
and summarized as a structure function in their Fig. 8, where
we see peaks of heightened brightness variability at time scales of
0.8 years and 9 years (in the quasar rest frame). The 0.8-year feature is
particularly well seen in their Fig. 14 for events in their lower
luminosity sample(Dark points) at a high level of significance. We consider
that structure function analysis is a poor way to study the 0.8-year
fluctuations, and that wavelet analysis, as applied to the Q0957 brightness
statistics by Schild (1999) would be more suitable.

The 9-year events are
considered somewhat uncertain by the De Vries et al because of problems
combining data sets related to different time scales. The structure
function for more rapid variability in the Q0957 quasar has already
been discussed in Section 3. Of course we now seek to match up the
De Vries et al (2005) structure function peaks with the fuel units,
and we start by associating the 0.8-year (300-day) events with
primordial fog particle fueling. De Vries et al (2005) stress that
these events are asymmetrical, having typically a slower rise time
and more rapid decay, and often but not always accompanied by a
color shift toward blue. Typical events of this kind can be seen in
the individual brightness curves from Giveon et al, who also reference
extensive earlier literature, and Hawkins (1996).

Typical events have amplitude 0.3 mag and time scales (in the quasar
rest frame) of approximately 300 days. These typical events have
also been seen in the Q0957 brightness record, where the (1+z)
cosmological redshift expands the expected 300-day time scale to
approximately 700 days. In the image A (less sensitive to
micro-lensing) brightness record shown as Fig. 1 in Pelt et al
(1998), three such events are seen peaking at Julian dates
JD - 2440000 = 5800, 7700, and 10,100. Observed amplitudes average
approximately 0.25 magnitudes. Thus these Q0957 peaks are typical of
unit fueling events seen in all quasars. However, further structure
in the Q0957 brightness history within one of these De Vries et al
``events'' has been interpreted by SV03 as resulting from internal
quasar structure on lags (observer's watch) 125, 190, 540, and 625
days. Thus for Q0957 the general peak in the De Vries structure
function curve includes some component from internal quasar
structure.

Since De Vries work implies that high-luminosity quasars vary less
than low-luminosity quasars, this result implies a quasar fueling
scenario in which: a) variations have a limited absolute magnitude
and, b) variations in luminosity are due to unit fueling events
involving sub-components or accretion instead of coherent variations
of the quasar (De Vries et al (2005)).

In this context we are now in a position to picture the quasar
response to a typical unit fueling event. The event has an
asymmetrical profile, with a rise time approximately twice as long
as the decay time. During the event, the luminous quasar output
increased approximately 30 \% in UV photons. Since the UV emission
is found at nearly the center of the quasar's output spectrum we
estimate from this that the quasar total output also increased by
approximately 30\%.

The average quasar luminosity fluctuation is observed to be on the
order of $1\times10^{44}$ ergs/sec, over a 300 day duration. If this
luminosity fluctuation is interpreted as being due to the integrated
response to a unit planetoid mass ingestion in the disk, then the
total integrated energy associated with this process is
$3\times10^{51}$ ergs.

Since the unit mass planetoid infall into the quasar accretion disk
occurs from large distances, the total energy which contributes to
the quasar disk luminosity on impact is equal to its relativistic
kinetic energy times an efficiency factor which is estimated to vary
between 1 and 10 percent. Since the rest mass energy of a
$1\times10^{-3} M_{\odot}$ primordial planetoid is comparable to the total
integrated energy of the 300 day quasar luminosity fluctuation, this
luminosity fluctuation event can be plausibly interpreted as being
caused by a collision of the quasar accretion disk with an infalling
planetoid with mass on the order of this size which is moving
relativistically. In this context the unit fueling process in
quasars can be explained energetically in terms of primordial
$1\times10^{-3} M_{\odot}$
planet-like particles infalling from large
distances where the energy conversion to UV luminosity does not need
to be very efficient.

This unit quasar fueling mass on the order of $1\times10^{-3} M_{\odot}$
is somewhat larger than the $1\times10^{-5} M\odot$ unit mass which was
previously inferred from quasar micro-lensing considerations (Schild,
1996). However we do not see this as a contradiction because the
micro-lensing process responds to the particles with the largest
optical depth, or to the peak of the mass distribution of the
micro-lensing particles which is on the order of  $1\times10^{-5} M_{\odot}$.
On the other hand the size of the unit mass ingestion associated
with the quasar luminosity fluctuation process will tend to select
the more massive $1\times10^{-5} M_{\odot}$
particles in the primordial
planetoid distribution, since the less massive in-falling particles
will cause smaller fluctuations not seen in the brightness records.

Since in our scenario the 300-day events found by De Vries et al
(2005) and earlier authors are attributable to planetary mass
ingestion events, this allows us to make a falsifiable prediction.
The structure function analysis presented by De Vries et al is not
optimal for studying the nature of the 300-day events because either
positive (brightening) or negative (fading) events could be
responsible. However our scenario predicts dominant positive events,
and a more optimum analysis with wavelets as demonstrated by Schild
(1999) would discriminate. Our unit fueling scenario predicts that
the 300-day events should be dominated by positive (brightening)
wavelets, and wavelet analysis would allow easily interpreted
discrimination.

\section{Radio Properties of Q0957 Measured By Micro-lensing}

In the following section, we analyze data for Fourier power of the A
and B quasar images and conclude that the B image shows more power
at micro-lensing frequencies. From this and calculation of the
coherence between the radio fluctuations seen, we are able to
conclude that the radio emitting region undergoes micro-lensing at
radio frequencies. When we estimate the observed amplitude of the
measured micro-lensing, we conclude that only a small fraction of
the observed radio brightness from the identified region is
micro-lensed in the observed "radio microlensing" events.

\subsection{The Coherence and Radio Brightness Imply Micro-lensing}

Radio emission at 6 cm has been monitored since the 1979 discovery
of the Q0957 quasar by the MIT radio group (Lehar et al, 1992;
Haarsma et al 1997, 1999). The purpose was to determine the time
delay, and one observation per month was made in general. At its
high ecliptic latitude, Q0957 +561 is far from the sun on any
calendar date, and the observing record is of high quality without
the annual dropouts characteristic of the optical data.

The basic process is revealed in the Fourier power spectrum for the
6 cm radio emission measured at the VLA. Comparison of the A and B
image power spectra shows extremely similar spectra from 0 - 3
cy/yr, which relates to variability on time scales longer than 120
days. But a factor 5 larger power level is seen for image B relative
to A between 3 and 5 cycles/yr. Since the radio flux was sampled
monthly, the sample frequency is 12 cy/yr and the Nyquist frequency
is 6 cy/yr, or 60 days. However, inspection of the brightness record
shows the existence of many brightness spikes indicated by 2 or more
observations measured with the 30-day basic sampling rate, and thus
many individual events having a barely resolved 60-day time scale.
Because the B image is seen through the lens galaxy its optical
depth to micro-lensing is higher than A. So if a stronger pattern of
fluctuations is seen, it is interpreted as arising in micro-lensing.
In other words, the A and B image are images of the same quasar, so
insofar as their brightness fluctuations differ, we attribute the
difference to the micro-lensing that originates in the lens galaxy,
and it would be strongest in image B which is seen through the lens
galaxy and has a $\times4$ higher optical depth to micro-lensing by
the granular structure of matter (stars, planets) in the lens
galaxy.

If the higher amplitude of fluctuations limited to 3.5 - 5 cy/yr is
caused by micro-lensing, then there must be more spatial structure
in the source or more micro-lensing particles of appropriate mass
for 3.5 - 5 cycles/yr. The Schild and Vakulik 2003 paper shows that
solar mass micro lenses can produce cusps on this time scale, for
the adopted standard transverse velocity of the cusp pattern.
However, for micro-lensing to occur, both the fine cusps due to
granularity of the mass distribution in the lens galaxy AND fine
structure in the radio emission region must be present. This allows
us to make size estimates for the radio emitting region. The
brightness fluctuations measured as power in Fourier spectral
estimates on time scales 3 - 5 cy/yr are best seen in the brightness
curves given in Haarsma et al (1999) Fig. 4. We interpret the solid
curves in the plot to show the smoothed brightness trend of the two
quasar images, and we find many pairs of points that reveal
brightness spikes relative to the smoothed trend line. These spikes
we interpret as micro-lensing events, and we find that they have
typical brightness amplitudes of 5 \% on the required time scale, 60
days. Typical events in the B image record are seen at JD - 2440000
= 6600, 7600, 7700, 9400, 9600, and perhaps 10,300. It may also be
seen that these events are more commonly seen in the B image than in
A, confirming the result of the Fourier power spectrum estimate
described above.

Again referring to the SV03 paper, at optical wavelengths it was
estimated from direct simulation that structure on the scale of
$2\times10^{16}$ cm produces fluctuations on time scales of 150
days. Thus we conclude from simple scaling arguments that the
emitting radio source has a radius of approximately
$2.2\times10^{16}$ cm for an adopted mean micro-lensing particle
mass of $0.5 M_\odot$ (Schild 1996) and for the adopted event
duration of 60 days.

\textbf{\subsection{The Micro-lensed Fraction of 6 cm Radio
Emission}}

For structure on such scales, the R/Ro test of Refsdal \& Stabell
(1991, 93, 97) cannot be applied because the size of the Einstein
ring for a half solar mass star, $2\times10^{16}$ cm, is comparable
to the size of the emitting region, as estimated above. Thus the
small amplitude of the observed 60-day brightness fluctuations is
not indicative of a very large source, but rather is indicative that
the micro-lensed source contains only a small fraction of the radio
luminosity, as we now estimate.

The amplitude of radio brightness microlensing fluctuations has not yet been
estimated by wavelet analysis as the optical has (Schild 1999).
However we easily estimate the amplitudes of some typical
cusp-shaped events seen in the published radio brightness history,
(Haarsma et al 1997, Fig. 4). Here we see that the
strongly microlensed B image has typical event amplitudes of 1.4
mJy relative to the mean 25 mJy flux, or approximately 5\%
micro-lensing amplitude. Similarly, the A image has 1.9 mJy
amplitude relative to a 35 mJy mean, or 5\% amplitude. Because
micro-lensing of a very compact source would produce events with
amplitudes a factor 10 or 20, we immediately conclude that the radio
source component with radius estimate $2.2 * 10^{16}$ cm contributes
only a small fraction, approximately a percent, of the total
measured 6 cm radio brightness.

It is further inferred that the remaining 99\% of the measured 6 cm
flux originates in a region too large to have micro-lensing events
on time scales of 30 days to 10 years, as has been previously
inferred. The rapid small brightness fluctuations have sometimes
been interpreted in the context of interstellar scintillations (Winn
et al, 2004; Koopmans et al, 2003). We consider that the conclusion that
over the 3 - 5 cy/yr micro-lensing frequency band, centered at
approximately 60 days, the Fourier power measured in the radio
brightness curves is higher in the B image than in the A image, in
approximately the same ratio as the micro-lensing probability, is a
strong indication that the fluctuations originate in micro-lensing.

The position of the micro-lensed radio core source along the jet
axis is also defined observationally. The radio and optical
brightness has been found to have a substantial coherence with a lag
of 35 days, with the optical preceding the radio (Thomson and Schild
1997). Thus apart from geometrical factors, the micro-lensed radio
emitting region lies $35/(1+z)$ light days above the inner quasar
accretion disk structure, or 14.5 light days $(3.6\times10^{16} cm)$
above the central source (in projection).
With correction for the calculated 54
degree inclination of the quasar's rotation axis with respect to the
line of sight, the compact micro-lensed radio emitting region is
$9\times10^{16}$ cm above the plane of the accretion disc. We
conclude that a micro-lensed compact region of radio emission,
contributing only a percent of the total radio flux, with a
$2\times10^{16}$ cm radius is located $9\times10^{16}$ cm above the
accretion disc plane.

\textbf{\subsection{Dynamic Origin of the Micro-Lensed Radio
Structure in Q0957+561 based on MECO Magnetic Field Line
Reconnection Processes}}

Recently the case of radio jet formation from a rotating collapsed
object with an intrinsic magnetic dipole field aligned along the
rotation axis has been studied (Romanova et al, 2004) In
that paper it was shown that such a
configuration would produce a tangled network of magnetic field
lines leading to the creation of magnetic towers associated with
vertical quasi-periodic outflows. For the case of the central MECO
in the quasar Q0957+561, we expect that a analogous type of
stretching, bunching, breaking, and reconnection of the MECO
magnetic field lines should occur at distances comparable to the
sizes of the interior quasar SV03 structure radio structure discussed above.

When such re-connection effects act on the MECO magnetic field
lines, the process is likely to entail their breaking and
re-connection at the local Alfven speed. Since the local Alfven
speed will be relativistic in close proximity to the rotating
central MECO object, we expect that the field line re-connection
process described above has the potential to create the relativistic
motions observed in radio jets.

The relativistic Alfven speed is
\begin{equation}
v_A^2=\frac{c^2B^2/4\pi}{\rho c^2+\gamma P/(\gamma-1) +B^2/4\pi}
\end{equation}
In the MECO model in the low hard state, hot plasma containing
magnetic field lines is quasi-periodically peeled off the inner
magnetospheric radius of the accretion disk in the form of a funnel
flow process similar to that described in Romanova et al (2004). In
the process of entering the magnetosphere, the poloidal magnetic
field components are wound toroidally until $B^2/4\pi \approx \rho
v_k^2$, where $v_k \sim 0.2c$ is the keplerian speed of the inner
disk.

At the inner disk radius $\gamma \sim 5/3$ and the gas pressure $P$,
which essentially matches the magnetic pressure of the poloidal
field component at the edge of the inner disk, is negligible by the
time the base of the jet is reached. For this reason, in the
toroidally wound donut that falls in through the magnetosphere, the
gas pressure $P$ is negligible. The end result of this fact is that
$B^2/4\pi \sim \rho c^2$ near the base of the jet.

The energy density required for plasma to escape from deep inside
the gravitational well is about $GM\rho/r_e$, where $r_e$ is the
"ejection radius". At this radius, the energy density of the wound
up toroidal magnetic field in the donut is $B_{\phi}^2/4\pi$ and
this is what drives the jet. Equating these energy densities leads
to $r_e \sim 4\pi G M \rho /B_{\phi}^2 = GM/v_A^2$. Taking $r_e =
2\times 10^{16}$ cm from the observed Schild-Vakulik structure,
leads to the result that $v_A = 0.16c$; hence the Alfven speed
associated with this process near the MECO is relativistic.
 
\textbf{\section{Theoretical Analysis and Overview of the Schild-Vakulik
Structure (SV03 structure) Inside of the Quasar Q0957+561}}

In this section we discuss the applicability of currently known
theoretical quasar models to the empirical Schild-Vakulik Structure
(SV03 structure) discussed in the preceding sections. Based upon the
observational evidence discussed, the quasar Q0957+561 has physical
elements as follows: (see figure 1): \textbf{(INSERT FIGURE
1)}
 
Elliptical Elvis Coronal Structure $R_e=2\times 10^{17}$ cm ,
$H_e=5\times
10^{16}$ cm\\
Inner Radius of Accretion Disk $R_m = 74 R_g = 3.9\times 10^{16}$
cm\\
Size of Hot Inner Accretion Disk Annulus $\delta(R_m) =
5.4\times
10^{14} cm = 1 R_g$ \\
Size and Location of Base of Radio Structure $R_r=2\times 10^{16}$ cm ,
$H_r=9\times 10^{16}$ cm \\
 
Note that some of the dimensions listed above related to the sizes of the
optical structures differ by factors of approximately 2 from the values
stated in SV03. This is because the projection deduced in Schild (2005)
implies projection factors that make the Fig. 1 sideways-viewed quantities
larger by about that factor. Our quoted dimensions in Table 2 (bottom)
correspond to the dimensions in Fig. 1.

The radial width of the UV-luminous Elvis structure, $R_{e}$ was estimated
in SV03 to be $2\times 10^{16}$ cm, and was estimated from microlensing of
the broad emission lines by Gordon et al (2004) to be a comparable
$1.4\times 10^{16}$ cm.

Since the quasar Q0957+561 has been
observed to contain a central compact object with a mass on the
order of $M = {(3 - 4)}\times10^{9}$ solar masses, the observed SV03
structure
inner accretion disk radius of $3.9\times 10^{16}$ cm implies that
the inner radius of the accretion disk is located at about 74 $R_g$.
In addition, at this radius a hot annulus of material exists with a
radial thickness of $5.4\times 10^{14}$ cm (which for the above mass
is on the order of 1 $R_g$ in radial thickness). Finally the
hyperbolic Elvis Coronal Structure which is observed in Q0957+561
appears, on the basis of its observed $H_r$ / $R_r$ ratio, to have a
rather wide opening angle of 76 degrees with respect to the z-axis
of rotation.

At first one might consider explaining the Schild-Vakulik structure
(SV03 structure) in terms of intrinsic magnetic moment generated by a
central
spinning charged black hole in the quasar Q0957+561. However this
explanation fails because the necessary charge on the spinning black
hole required to make it work would not be stable enough to account
for the long lifetime of the SV03 structure . This inherent instability occurs
because the value of the charge/mass ratios of electrons and protons
implies that opposing electric forces on them would then be at least
an order of magnitude larger that their gravitational attraction to
the central black hole. In addition it would be difficult to
maintain this required charge on the spinning black hole if it were
also surrounded by an accretion disk like that observed in quasar
Q0957.

Since Q0957+561 appears to be in a radio loud Low Hard State with a
very large 74 $R_g$ inner accretion disk radius a second attempt to
explain the above observations would be to use a standard Kerr Black
Hole-ADAF-Accretion Disk-Jet Model (Narayan and Quataert, 2005).
One would do this by choosing
the parameters in this scenario so that the transition from the hot
thin outer disk to the inner ADAF disk occurred at the 74 $R_g$
radius. However since the magnetic field in this model is anchored
to the accretion disk and not to the central Kerr Black Hole
this model cannot account for either: a) the very thin 1 $R_g$
radial size of the observed Hot Inner Accretion Disk Annulus at the
74 $R_g$ radius or, b) the wide opening Elvis outflow opening angle
of 76 degrees with respect to the z-axis of rotation (since the
frame dragging on the disk magnetic field that occurs in Kerr Black
Hole-ADAF-Accretion Disk-Jet Models characteristically generates
jet-like outflows with relatively narrow opening angles on the order
of 30 degrees with respect to the z-axis of rotation, e.g. see
figure 1 of the Narayan and Quataert article quoted above).

Since the standard Black Hole-ADAF-Accretion Disk-Jet Model model
does not predict the SV03 structure correctly, a third and final attempt to
find a black hole description for the SV03 structure observed in Q0957+561
would be to apply the Magnetically Arrested Disk (MAD) black hole
model (Igumenschev, Narayan, and Abramowicz, 2003), to the above
described Q0957+561 structure. In this case MAD parameters
would be chosen which would set the magnetospheric radius of the MAD
accretion disk at 74 $R_g$. However in this MAD model description of
Q0957+561, instead of the observed hot annular band of material of
radial thickness on the order of 1 $R_g$, there would now occur at
this radius a stochastic injection of hot blobs of plasma which
would orbit into the black hole while emitting visible radiation
from 74 $R_g$ all the way down to the photon orbit at 3 R$_g$. Based
on the Keplerian orbital periods at 74 R$_g$, the MAD-black hole
model would then generate observable periodic luminosity
fluctuations whose local periods at the quasar, over the radial interval
ranging from 74 Rg down to the photon orbit at 3 Rg, were on the
order of 800 days. Hence
if this MAD-Black Hole model had been operating in Q0957, the clean
fluctuation signal associated with the inner hot ring structure of
the Schild-Vakulik model would not have been seen because of the
smearing out of all of the MAD periodic QPO that would be occurring
in the orbital range from 74 $R_g$ down to 3 $R_g$. Furthermore,
periodic
brightness fluctuations produced by luminous hot blobs crossing caustics
created by the complex mass distribution in lens galaxy G1 would produce
highly periodic, approximately 100-day, brightness spikes not observed.

Hence we find that neither spinning charged black holes, the
standard Black Hole-ADAF-Accretion Disk-Jet Model, nor the Black
Hole Magnetically Arrested Accretion Disk (MAD) Black hole model,
(the latter two of which contain magnetic fields that are assumed to
be anchored in the accretion disk and not anchored to the central
compact black hole object) can account for all four of the
components of SV03 structure observed within the quasar Q0957+561

Having tried all of the plausible black hole models and found them
unsatisfactory we will now show that the four components of the
Schild-Vakulik structure in the quasar Q0957+561 can be consistently
described within the context of the Magnetospheric Eternally
Collapsing Object (MECO) model, (Robertson and Leiter, 2002, 2003,
2004, 2005;
Leiter and Robertson, 2003),
in which a very strong intrinsic magnetic field, anchored
to a slowly rotating highly redshifted central compact MECO object,
interacts in a magnetic propeller mode with the surrounding
accretion disk and generates the four components of the SV03 structure
with
Elvis coronal outflows that can have a wide opening angle greater that
60 degrees with respect to the z-axis of rotation.

In the following discussion the formal details about the MECO model
which appear in the above papers have been summarized into a series of short
appendices (1-11) for the convenience of the reader and will be referred
to as needed in order to clarify the details of specific calculations.
More specific information and simulations for the magnetic propeller
mode of central compact magnetic objects surrounded by accretion disks can be found in
Romanova et al, (2002, 2003a).

We begin our MECO analysis of the SV03 structure in Q0957+561 starting
with Table 1 below in which: a) the left hand column summarizes the MECO
physical quantities relevant to Q0957+561 and the location in the
appendicies where for the convenience of the reader the associated
derivations and discussions about these physical quantities can be
found; b) the middle column gives the specific form of the equations
to be used in the calculation of these physical quantities and; c)
the right hand column gives the functional form of the relevant mass
scaling of these physical quantities in order to demonstrate clearly
how the equations for these physical quantities can be applied to
the case of both galactic black hole candidates (GBHC) and AGN.

Then, assuming that: a) the MECO behavior of this quasar in the
radio loud Low Hard State is similar to that of an average MECO-GBHC
in the radio loud Low Hard State (like that analyzed in the
Robertson-Leiter papers discussed above), and b) using the observed
values of: X-ray luminosity, radio luminosity, red shift (z=1.4),
and the physical dimensions of the internal SV03 structure for Q0957+561
as
input in the mass scaled MECO equations in Table 1, then c) the
results shown in Table 2 are obtained.

Table 2 is intended to be read as follows: in the upper part, the scaling
relations from Table 1 and other input data are used to determine the
properties
of the GBHC and AGN. In the lower panel, the empirical size scale
parameters,
determined by Schild and Vakulik and corrected for the specific quasar
orientation factors in Schild (2005), are scaled and tabulated for the GBHC
case according to our adopted scaling relations.

In this manner we found that the presence of a $3.6\times10^{9}$
solar mass MECO acting as the central compact object in the quasar
Q0957+561 consistently predicted all of the four components of the
observed SV03 structure. The physical picture which emerged was as
follows: 1)
since Q0957 is a radio loud quasar in a LHS then the inner region of
the MECO accretion disk at the magnetospheric radius $R_m$ is larger
than the co-rotation radius $R_c$ and the non-thermal MECO-LHS X-ray
emission will be generated by the magnetic propeller interaction of
the intrinsic magnetic moment in the central MECO with the inner
accretion disk. 2)The "line-driving" force acting on atoms within the
radiation field environment around the MECO, working in conjunction with
the MECO intrinsic magnetic propellor interaction with the
accretion disc and the magnetic corona about. acts to generates the
hyperbolic Elvis outflow of plasma $2\times10^{17}$ cm from the central
object and with the wide opening angle observed, leading to the development 
of a radio jet. This radio emitting region
has radius $R_r$ given by $2\times10^{16}$ cm 
and height $H_r$ of $9\times10^{16}$ cm. Non-themal continuum emission also
reveals an Elvis coronal structure
of radius $2\times10^{17}$ cm above the accretion disc plane (Fig. 1).
where the MECO magnetic propeller driven ionized matter
is reprocessed into non-thermal UV and Optical emissions.
3) The inner region of the accretion disk will be located at $R_m$ =
1.4 $R_c$ where $R_m$ is the magnetospheric radius, and the MECO
co-rotation radius $R_c$ is 53 $R_g$. A high density,
optically thick, radially thin inner band of material of radial
thickness given by 0.014 $R_m$ and equal to $5.4\times10^{14}$
cm is formed by the MECO magnetic propeller at the magnetospheric
radius $R_m$. The thermal emission from this hot optically thick
inner band of radial thickness Rm then accounts for the observed
small blue thermal UV bump and its micro-lensing fluctuations.
Normalizing the optically thick MECO accretion disk to that of a
NS-XRB, we find that the temperature $T_t$ of the thermal emissions
from the thin hot inner annulus of material of radial thickness
$R_m$ is on the order or 7200 K. For the observed Q0957 red shift of
$z =1.4$ this thin hot inner annulus generates the observed local
thermal "little blue bump" emission at wavelength 
2700 Angstroms with a thermal UV luminosity $L_t$ equal to
$1.4\times10^{44}$ erg/sec (which is equal to about one quarter of
the total non-thermal UV luminosity which is generated by Compton
up-scattering from the larger hyperbolic Elvis Coronal Structure,
also present)
\begin{table*}
\scriptsize
\begin{center}
\caption{MASS SCALING EQUATIONS ASSOCIATED WITH VERY
HIGH REDSHIFT EDDINGTON LIMITED MECO [REF APPENDIX 1-11]}
\end{center}
\begin{tabular}{lll} \hline
MECO Physical Quantity & Equation & ($M=Mass/M_\odot$)\\
\hline
Surface Redshift - (App 1, 9-10) & $1+z_s = 1.5\times 10^8
(M/7)^{1/2}$
&
$M^{1/2}~$\\
Surface Luminosity Obs.- (App 5) & $L_s=L_{Edd}/(1+z_s)$ erg/s &
$M^{1/2}~$
\\
Surface Temp Obs. - (App 5) & $T_s=2.3\times 10^7/[M(1+z_s)]^{1/4}$
K &
$M^{-
3/8}$\\
Rotation Rate - (App 7) & $\nu_2=0.89[L_{q,32}/M]^{0.763}/L_{c,36}$
Hz &
$M^{-
1}~$\\
Quiescent Lum. - (App 7) & $L_{q,32}=0.65 M [L_{c,36}/M]^{1.75}$
erg/s &
$M$~~\\
Co-rotation Radius - (App 7)& $R_c=46.7 R_g/[M \nu_2]^{2/3}$ cm &
$M$~~\\
Magnetosphere Radius - (App 7) & $R_m= 3.33
R_g[\mu_{27}^4/((L_x/L_{Edd})^2 M^{10})]^{1/7}$ cm &
$M$~~\\
Magnetic Moment - (App 7)& $\mu_{27} = 8.16[L_{c,36}
M/\nu_2^3]^{1/2}$ cm&
$M^{5/2}~$\\
Magnetic Field - (App 7)& $B_{\theta,10} = 1.12 (R_g/r)^3 [L_{c,36}/(M^5
\nu_2^3)]^{1/2}$ gauss
& $M^{-1/2}$\\
Radio Luminosity - (App 8)& $L_{rad,36}=10^{-6.635}M^{0.835}
L_{x,36}^{2/3}[1-(L_{x,36}/L_{c,36})^{1/3}]$ erg/s & $M^{3/2}$~ \\
\hline
\end{tabular}\\
\bigskip
\begin{center}
\caption {MASS SCALING OF LHS MECO-GBHC TO QUASAR Q0957}
\end{center}
\begin{tabular}{lll} \hline
MECO Physical Quantity & AVG. MECO-GBHC & MECO-QUASAR
Q0957\\
\hline
Mass of Central MECO & $7$ $M_\odot$ & $3.6\times 10^9 M_\odot$\\
Surface Redshift & $1.5\times 10^8$ & $3.4\times 10^{12}$ \\
Surface Luminosity Obs. & $6.1\times 10^{30}$ erg/sec & $1.4\times
10^{35}$
erg/sec\\
Surface Temp Obs. & $1.3\times 10^5$ K & 69 K\\
Rotation Rate & $12$ Hz & $2.3\times 10^{-8}$ Hz\\
Intrinsic Mag. Moment & $2.1\times 10^{30}$ gauss-cm$^3$ & $1.2\times
10^{52}$
gauss-cm$^3$\\
Magnetic Field Obs. & $(8.3\times 10^9$ gauss)($6Rg / r)^3$ &
$(3.6\times
10^5$
gauss) $(6Rg / r)^3$\\
Quiescent X-Ray Lum. & $2\times 10^{33}$ erg/sec & $9.4 \times
10^{41}$
erg/sec\\
Co-rotation Radius & $53 R_g$ & $53 R_g$\\
X-Ray Lum. & $4.7 \times 10^{36}$ erg/sec & $2.4\times 10^{45}$
erg/sec\\
Radio Lum. & $1.1\times 10^{30}$ erg/sec & $1.3\times 10^{43}$
erg/sec\\
\\
\hline
Schild-Vakulik Structure & AVG. MECO-GBHC & MECO-QUASAR
Q0957\\
\hline Magnetosphere Radius & $R_m = 74 R_g = 7.6\times 10^7$ cm
& $R_m = 74 R_g = 4\times 10^{16}$ cm\\
Hot Inner Disk Band & $\delta(R) = R_g = 10^6$ cm & $\delta(R) = Rg
=
5.4\times 10^{14}$ cm\\
Base of Radio Jet & $R_r=5\times10^7$ cm $H=2\times10^8$ cm &
$R_r=2\times10^{16}$ cm $H=9\times10^{16}$ cm \\
Elvis Corona$^a$ & $R_e=5\times10^8$ cm $H=1.2\times10^8$ cm
&$R_e=2\times10^{17}$ cm $H=5\times10^{16}$ cm\\
\hline
\end{tabular}\\
\bigskip
$^a$Elvis Coronal Structure may not be seen in hot LHS MECO-GBHC
accretion
disks.\\
\end{table*}
 
\newpage
Hence in the context of the MECO model we have shown that the quasar
Q0957+561 in a radio loud Low Hard State is physically similar to
a mass scaled up version of an average MECO-GBHC where,
because of the cooler accretion disk which occurs in the quasar, all
four components of the SV03 structure are observable. The basic
observable
elements of the SV03 structure in this quasar are: 1) a large Elvis elliptical
coronal structure with a 76 degree opening angle with respect to the
z-axis of rotation; 2) a large magneto-spherically generated inner
accretion disk radius; 3) a thin hot inner accretion disk annulus
located at the magnetospheric radius; and 4) a radio structure whose
base is located directly above the central MECO

We have demonstrated that the MECO model is able to explain all of
the properties of the SV03 structure observed in the quasar Q0957+561 as
intrinsic magnetic propeller interactions with the accretion disk.
Hence on this basis we conclude that the presence of the SV03 structure
represents observable evidence for the existence of an intrinsic
magnetic moment generated by a central MECO in the heart of this
quasar.
\newpage
\textbf{\section{Conclusions}}
We have examined the empirical data for the lensed and micro-lensed
Q0957+561 A,B quasar obtained from 20 years of brightness monitoring
at visible wavelengths (near-ultraviolet emission at the quasar). We
have also examined several conclusions inferred previously from
analysis of the auto-correlation and micro-lensing properties of the
monitoring data, and we have now collected these results in a
consistent presentation that can confront physical quasar models and
their simulations.

The structures and luminosities found, referred to as the
Schild-Vakulik structure (SV03 structure), are associated with a bright
inner edge of the accretion disc, surrounded by a coronal outflow (Elvis
structure) long known to explain the complex spectroscopic behavior
observed in quasars. However it is observed that the opening angle
of the coronal Elvis structure with respect to the z-axis of
rotation appears to have a very large value of 76 degrees. In
addition a radio emission region has been located directly above the
compact source. In particular the size and location of this radio
emitting region has been found to be where the reconnection
of magnetic field lines at relativistic Alfven speeds, like that
generated by a rotating central object containing an intrinsic
magnetic field, should occur.

Although our data do not include near-infrared data related to the dusty
torus, the standard picture now shows that this torus structure lies 20 \%
farther from the central source than the Elvis outflow structures, and is
thus presumably in a velocity dead zone shielded by the Elvis outflow.
Notice that the Suganuma et al (2004) reference shows in their
Table 2 that for Seyfert Galaxy NGC 5548, the highest excitation lines of
Hydrogen and O III lie closest to the central source, with the low
excitation lines of O I and Mg II twice as far out, and the K-band
infrared emission 20 \% further out still. It is difficult to imagine that
the infrared emitting dust particles could survive the intense X-ray and
ultraviolet fluxes originating at the central source, and it is most likely
that the dusty torus lies in the equatorial region sheltered by the Elvis
outflow structures.

In the discussion in section 7 we showed that attempts to model
observation of the SV03 structure in the quasar Q0957+561 in terms of the
intrinsic magnetic moment generated by a central spinning charged
black hole in the quasar Q0957+561 failed because the necessary
charge on the spinning black hole required to make it work would not
be stable enough to account for the long lifetime of the SV03 structure.
Similarly attempts to model the SV03 structure in terms of the class of Kerr
Black Hole-ADAF-Accretion Disc Corona-Jet Models, in which the
magnetic field is intrinsic to the accretion disk and not the
central rotating black hole, were shown to have difficulty in that
they were unable to account for the very large opening angles which
are observed for the coronal Elvis outflows. Finally we showed that
Magnetically Arrested Disc (MAD) black hole models also have
problems in that they predict the existence of orbiting infalling
hot blobs of plasma inside of the inner region of the accretion disk
that are not observed.

On the other hand we have found that the SV03 empirical structure can
be successfully explained by Magnetospheric Eternally Collapsing
Object (MECO) models, which feature a highly redshifted, Eddington
limited, collapsing central compact object containing a strong
intrinsic magnetic field aligned with the MECO axis of rotation. In
this model the resultant MECO magnetic propeller effects which
interact with the inner regions of the accretion disc, create an
inner luminous annular (band like) structure and an outer coronal
structure characterized by strong relativistic outflow with a wide
opening angle to the z-axis of rotation as is observed in the SV03
empirical structure. In particular the size and location of the
radio emitting region associated with the SV03 structure in the quasar
Q0957+561 has been found to be located in the region above the
central compact object where the reconnection of magnetic field
lines at relativistic Alfven speeds, like that generated by a
rotating central MECO containing an intrinsic magnetic field, should
occur

The MECO contains a central rotating magnetic object whose dynamo
sweeps clean the central region of the quasar out to a distance
where the magnetic propeller acts on the inner edge of the
accretion disc, and a radio emitting region above the disc where
magnetic field line must twist and bunch up until they eventually
break and reconnect at relativistic speeds. Such an object does not
have an event horizon; instead, in-falling material collects at an
inner structure just beyond 2$R_g$ that further collapses to higher
red shift while remaining in causal connection for all time. Because
of the small light cone angle for radiation escaping from this
highly redshifted region to the distant observer, the resultant low
luminosity in the far-infrared wavelengths make this region
difficult to detect in the case of quasar Q0957+561.

It is important to note that the MECO model that seems to best fit
the empirical SV03 structure in the quasar Q0957+561 differs
significantly from most black hole models currently under
consideration. In particular the predicted SV03 empirical structures
they generate seem to resemble the complex inflow-outflow pattern
seen in magnetic propeller models for young stellar objects. The
action of such magnetic propeller forces has been discussed and
simulated by Romanova et al (2003a,b, 2004) with non-relativistic
models that produce observable structures that are closely similar
to the observed Schild-Vakulik structure.

We have also found that the MECO explanation of the empirical SV03
structure implies an interesting similarity between the MECO model
for the quasar Q0957+561 and the MECO models that explain the
phenomena associated with Galactic Black Hole Candidates in the Low
Hard state with luminosity on the order of one tenth of a percent of
the Eddington limit luminosity.

On the basis of the above observational and theoretical arguments we
conclude that the observation of the Schild-Vakulik structure
in the quasar Q0957+561 represents strong evidence for the existence
of an observable intrinsic magnetic moment, generated by a
supermassive $3.6\times 10^9 M_\odot$ MECO acting as the central
compact object in this active galaxy, which implies that this quasar
does not have an event horizon.
\section{Appendix 1 - Magnetospheric Eternally Collapsing Objects
(MECO)}
Using the Einstein-Maxwell Equations and Quantum Electrodynamics in
the context of General Relativistic plasma astrophysics we will show
that it is possible to virtually stop and maintain a slow, steady
(many Hubble times!) secular collapse of a compact physical plasma
object outside of its Schwarzschild radius with photon pressure
generated by synchrotron radiation from an equipartition surface
magnetic field. To control the rate of collapse, the object must
radiate at the local Eddington limit, but from a highly redshifted
surface.

In Appendix 9 and 10 of this paper we will show that surface drift
currents within a pair plasma at the MECO surface can be shown to be
able to generate the required fields. Drift currents proportional to
${\bf g\times B}/B^2$ occur for plasmas at rest in gravitational and
magnetic fields. The equatorial poloidal magnetic field associated
with an Eddington limited secular rate of collapse of the highly
redshifted exterior surface of the MECO can be shown to be
controlled by quantum electrodynamic pair production processes in
strong magnetic fields to be on the order of $\sim 10^{20}~gauss$.
Magnetic fields of this magnitude are strong enough to create a
redshift dependent source of bound electron-positron pairs in the
plasma at the surface of the MECO (in addition to the copious
production of electron-positron pairs created by photon-photon
collisions in the highly compact pair dominated plasma in the MECO),
whose interaction with the Eddington limited synchroton radiation
acts to stabilize the secular collapse rate. The magnetic field of
the interior is approximately what one would expect from flux
compression during collapse, $\sim 2.5\times
10^{13}\sqrt{7M_\odot/M}$ gauss and its radial component is
continuous across the surface boundary. The poloidal field is
discontinuous across the surface and much stronger externally due to
the surface drift currents. We will show in Appendix 9 that at the
MECO surface radius Rs (which is slightly larger than the
Schwarzschild radius $2GM/c^2$), the ratio of poloidal field on the
surface to the poloidal field just under the MECO surface is given
by
\begin{equation}
B_{\theta,S^+}/B_{\theta,S^-}=(1+z_s)/(2ln(1+z_s))=10^{20}/(2.5\times
10^{13})\sqrt{7M_\odot/M})
\end{equation}
where $z_s$ is the surface redshift. This has the solution
\begin{equation}
1+z_s = 1.5\times 10^8\sqrt{M/7M_\odot}
\end{equation}
The $\sim 10^{20}~gauss$ poloidal magnetic field on the surface of
the MECO generates an intrinsic magnetic moment which when distantly
observed is reduced, by a factor of $3(1+z_s) =4.5\times
10^8\sqrt{M/7M_\odot}$ due to the surface redshift of the MECO, to a
level which agrees well with the intrinsic magnetic moments observed
in GBHC and AGN. The surface luminosity is also reduced below the
conventional Newtonian Eddington limit by $(1+z_s)$ when distantly
observed, and the decay lifetime is extended by the same factor.
Figure 2 represents a schematic diagram summarizing the physical
mechanisms contributing to the creation of a MECO in the general
relativistic gravitational collapse process, which are discussed in
the appendix 2-11 sections of this paper. (\textbf{INSERT FIGURE 2)}

\section{Appendix 2 - The Strong Principle of Equivalence}

Astrophysicists nowadays generally accept the inevitability of the
curvature singularities of black holes \footnote {The modern notion
of a black hole began with Hilbert's error of application of
boundary conditions for the solutions of Einstein's field equations
of General Relativity for a mass point. Hilbert's solution has been
erroneously attributed to Schwarzschild, however Schwarzschild's
preceding original solution had no event horizon. [see Abrams 1979,
1989] Nevertheless, Hilbert's solution is analytically extendible
through the horizon to a central singularity, hence the modern black
hole. The fact that the radius of the event horizon, which is
directly proportional to the gravitational mass, can be changed
arbitrarily by (generally singular) coordinate transformation
strongly suggests that the horizon is unphysical.}, however, if GBHC
and AGN are observationally confirmed as containing intrinsic
magnetic moments, this will be nature's way of telling us that such
singularities are not really permitted to exist. For black holes to
exist, gravity must be able to do what no other force of nature can
do; namely, to accelerate finite mass to exactly the speed of light.
But this means that horizon crossing geodesics would become null
rather than timelike. In General Relativity (GR) the Strong
Principle of Equivalence (SPE) requires that Special Relativity (SR)
must hold locally for all time-like observers in all of spacetime.
This SPE requirement is a tensor relationship that implies that (i)
the spacetime manifold for observers located in field-free regions,
distant from gravitating masses, must approach the flat spacetime of
SR \footnote{This eliminates Kruskal-Szekeres coordinates from the
real world of astrophysics.} and (ii) the spacetime world lines of
massive matter must always be timelike. \footnote{
Models of gravitational collapse that lead to the
development of event horizons and central curvature singularities
inevitably abandon the SPOE requirement for timelike world line
completeness. The vanishing of the metric time coefficient,
$g_{tt}$, at the Schwarzschild radius is sufficient to cause the
timelike physical three-speed of particles in radial free fall to
approach the speed of light there. In obvious notation, [Landau \&
Lifshitz 1975]. $V^2 = (\frac{dl}{d\tau_s})^2 =
c^2\frac{({g_{tr}g_{tr}} - g_{rr}g_{tt})v^r v^r}
{(g_{tt} + g_{tr} v^r)^2}$ where $v^r=\frac{dr}{d\tau}$.
In non-singular Finkelstein or Kerr-Schild coordinates, for which
$g_{tr} \neq 0$, we we find $V \rightarrow c$ as $g_{tt}
\rightarrow 0$. It has been shown [Leiter \& Robertson 2003] that
$ds^2 \rightarrow 0$ at surfaces of infinite redshift. In
Kruskal-Szekeres coordinates, in which $g_{tt}$ does not vanish,
there is no surface of infinite redshift at the Schwarzschild
radius, $R=2GM/c^2$, and timelike test particle geodesics can
traverse it \textit{in either direction}, so long as the initial
conditions are chosen in a manner that permits the `time'
coordinate to change in a positive sense. However, a central
singularity still exists in these coordinates and they have a
surface of infinite redshift as $r \rightarrow \infty$, at which
$ds^2 \rightarrow 0 $. This is an extreme example of a coordinate
transformation changing the size of the event horizon radius.}
Such spacetime manifolds are known
as `bundle complete' [Wheeler \& Ciuofolini 1995].
As a guiding principle, we look for solutions of the Einstein
equations
\begin{equation}
G^{\mu\nu}=(8\pi G/c^4) T^{\mu\nu}
\end{equation}
that satisfy the SPE requirement for timelike world line
completeness. Since there is nothing in the Einstein tensor
$G^{\mu\nu}$ that enforces this condition, we must rely on
non-gravitational forces in $T^{\mu \nu}$ to dynamically enforce it.
Since the energy-momentum tensor $T^{\mu\nu}$ serves as both a
source of curvature in the Einstein equations and a generator of the
equations of motion of matter, constraints on $T^{\mu\nu}$ that
enforce timelike world line completeness can also eliminate the
occurrence of event horizons. Thus the SPE requires that the right
hand side of the GR field equation must contain non-gravitational
elements capable of stopping the collapse of physical matter before
the formation of a `trapped surface'. This dynamically escapes the
Hawking and Penrose theorem which states that once a trapped surface
is formed, an event horizon and curvature singularities are
unavoidable.

In what follows we will show how the SPE constrains the solutions of
the Einstein field equations. Consider a comoving interior metric
given by
\begin{equation}
ds^2=A(r,t)^2c^2dt^2 - B(r,t)^2dr^2- R(r,t)^2(d\theta^2 +
sin^2\theta d\phi^2)
\end{equation}
and an exterior Vaidya metric with outgoing radiation (For a more detailed
discussion of the physical implications associated with our use of the
exterior Vaidya metric instead of the Kerr-Schild metric used by most
theorists, see Appendix 11)
\begin{equation}
ds^2=(1-2GM/c^2R)c^2du^2 + 2c du dR - R^2(d\theta^2 + sin^2\theta
d\phi^2)
\end{equation}
where $R$ is the areal radius and $u=t-R/c$ is the retarded
observer time. Following Lindquist, Schwarz \& Misner [1965], we
define
\begin{equation}
\Gamma= \frac{dR}{dl}
\end{equation}
\begin{equation}
U= \frac{dR}{d\tau}
\end{equation}
\begin{equation}
M(r,t)= 4\pi \int_{0}^{r}{\rho R^2 \frac{dR}{dr} {dr}}
\end{equation}
\begin{equation}
\Gamma^2 = (\frac{dR}{dl})^2 = 1 - \frac{2GM(r,t)}{c^2R} +
\frac{U^2}{c^2}
\end{equation}
where $dl$ is a proper length element in a zero angular momentum
comoving frame, $d\tau$ an increment of proper time, $U$ is the
proper time rate of change of the radius associated with the
invariant circumference of the collapsing mass, and $M(r,t)$ is
the mass enclosed within this radius. The last two of the
relations above have been obtained from the $G_0^0$ component of
the field equation [Lindquist, Schwarz \& Misner 1965]. At the
boundary of the collapsing, radiating surface, s, we find that the
proper time will be positive definite, as required for timelike
world line completeness if
\begin{equation}
d\tau_s= \frac{du}{1+z_s} = du((1 - \frac{2GM(r,t)_s}{c^2R_s} +
\frac{U_s^2}{c^2})^{1/2} + \frac{U_s}{c}) > 0
\end{equation}
where $z_s$ is the distantly observed redshift of the collapsing
surface. From the above equation we see that in order to avoid a
violation of the requirement of timelike world line completenes for
$Us < 0$, it is necessary to dynamically enforce the `no trapped
surface condition'. \footnote {It might be argued that there might
not be a surface that physically divides matter from radiation
inside a collapsing massive continuum, however, it has been shown
[Mitra 2000, 2002, Leiter \& Robertson 2003] that the above
equations and the $G_0^0$ field equation in a zero angular momentum
comoving frame produces the `no trapped surface condition' for any
interior R(r,t), provided that B(r,t) does not become singular at a
location where A(r,t) vanishes. However this requirement will be
satisfied as long as timelike world line completeness is maintained
by photon pressure generated by the equipartition magnetic field
everywhere in the comoving frame. We can consider any interior
location and the radiation flux there without requiring a joined
Vaidya metric. But there will ultimately be an outer radiating
boundary and the required match to the non-singular outgoing
exterior Vaidya metric guarantees that there will be no metric
singularity there.}
\begin{equation}
\frac{2GM_s}{c^2R_s} < 1
\end{equation}

\section{Appendix 3 - A Radiating, Collapsing, Magnetic Object}

The simplest form of the energy-momentum tensor that can satisfy the
SPE requirement of time like world line completeness, is one that
describes a collapsing, radiating plasma with an equipartition
magnetic field that emits outgoing radiation. Between the extremes
of pure magnetic energy [Thorne 1965] and weakly magnetic, radiation
dominated polytropic gases or pressureless dust [Baumgarte \&
Shapiro 2003] there are cases where the rate of collapse can be
stable. To first order, in an Eddington limited radiation dominated
context, these can be described by the energy momentum tensor:
\begin{equation}
T_{\mu}^{\nu} = (\rho + P/c^2)u_\mu u^\nu - P \delta_\mu^\nu +
E_\mu^\nu
\end{equation}
where $E_\mu^\nu = Qk_\mu k^\nu$, $k_\mu k^\mu = 0$ describes
outgoing radiation in a geometric optics approximation, $\rho$ is
energy density of matter and $P$ the pressure.\footnote {Energy
momentum tensors corresponding to metrics describing ingoing
radiation, which are used in many black hole model calculations,
(e.g. Baumgarte \& Shapiro [2003]) cannot be used here because
they are incompatible with the $Q > 0$ boundary conditions
associated with collapsing, outwardly radiating objects.} Here $Q$
is given by
\begin{equation}
Q=\frac{-(dM/du)/4 \pi R^2}{(\Gamma_s + U_s/c)^2}
\end{equation}
At the comoving MECO surface the luminosity is
\begin{equation}
L= 4\pi R^2 Q~~ >0.
\end{equation}
and the distantly observed luminosity is
\begin{equation}
L_{\infty} = -c^2\frac{dM_s}{du} = -c^2\frac{dM_s}{d\tau(1 + z_s)}
\end{equation}
After examining the relations between surface and distantly
observed luminosities, we will use this relation to determine the
MECO lifetime.
\section{Appendix 4 - Eddington limited MECO}
Among the various equations associated with the collapse
process there are three proper time differential equations applicable to
a compact collapsing and radiating physical surface. When evaluated on the
physical surface [Hernandez Jr.\& Misner, 1966,
Lindquist, Schwartz \& Misner 1965, Misner 1965, Lindquist,
1966] these equations are:
\begin{equation}
\frac{dU_s}{d\tau} = (\frac{\Gamma^2}{\rho+P/c^2})_s
(-\frac{\partial P}{\partial R})_s - (\frac{G(M + 4\pi R^3 (P + Q
) / c^2)}{R^2})_s
\end{equation}
\begin{equation}
\frac{dM_s}{d\tau} = - (4\pi R^2 P c \frac{U}{c})_s - (L (\frac{U}{c} +
\Gamma))_s
\end{equation}
\begin{equation}
\frac{d\Gamma_s}{d\tau} = \frac{G}{c^4}(\frac{L}{R})_s +
\frac{U_s}{c^2} (\frac{\Gamma^2}{\rho+ P /c^2})_s (-\frac{\partial
P}{\partial R})_s
\end{equation}
In Eddington limited steady collapse, the conditions $dU_s/d\tau
=0$ and $U_s \approx 0$ hold after some time, $\tau_{Edd}$, that
has elaspsed in reaching the Eddington limited state. Then
\begin{equation}
\frac{dU_s}{d\tau} = \frac{\Gamma_s^2}{(\rho +
P/c^2)_s}(-\frac{\partial P}{\partial R})_s
- \frac{GM_s}{R_s^2} = 0
\end{equation}
Where
\begin{equation}
M_s = (M + 4\pi R^3(P + Q )/c^2)_s
\end{equation}
includes the magnetic field energy in the pressure term and radiant energy
in
Q.
Eq. (19) when integrated over a closed surface can be solved for the
net outward flow of Eddington limited luminosity through the
surface. Taking the escape cone factor of $27(R_g/R_s)^2/(1+z_s)^2$
into account, where $R_g=GM/c^2$, the outflowing (but not all
escaping) surface luminosity, L, would be
\begin{equation}
L_{Edd}(outflow)_s =\frac{4\pi G M_s c R^2(1 + z_{Edd})^3}
{27 \kappa R_g^2}
\end{equation}
where $\kappa \approx 0.4$ cm$^2 / g$ is the plasma opacity. (For
simplicity, we have assumed here that the luminosity actually
escapes from the MECO surface rather than after conveyance through a
MECO atmosphere and photosphere. The end result is the same for
distant observers.) However the luminosity $L_s$ which appears in
the above equations is actually the net luminosity, which escapes
through the photon sphere, and is given by $L_s = L_{Edd}(escape)_s
= L_{Edd}(outflow) - L_{Edd}(fallback) =
L_{Edd,s}-L_{Edd,s}(1-27R_g^2/(R(1+z_{Edd}))^2$ Thus in Eq.s (17)
and (18), the $L_s$ appearing there is given by
\begin{equation}
L_s= L_{Edd}(escape)_s = \frac{4\pi GM(\tau)_s c
(1+z_{Edd})}{\kappa}
\end{equation}
In this context from Eq.s (9), (10), (17) and (22) we have
\begin{equation}
c^2\frac{dM_s}{d\tau} = -\frac{L_{Edd}(escape)_s}{(1+z_s)}
= - \frac{4\pi G M(\tau)_s c}{\kappa}
\end{equation}
which can be integrated to give
\begin{equation}
M_s(\tau) = M_s(\tau_{Edd}) \exp{((-4\pi G / \kappa c)(\tau -
\tau_{Edd}))}
\end{equation}
This yields a distantly observed MECO lifetime of $(1+z_s)\kappa
c/4\pi G \sim 5\times 10^{16}$ yr for $z_s \sim 10^8$. Finally,
equation (18) becomes
\begin{equation}
\frac{d\Gamma_s}{d\tau}
=\frac{G}{c^4}\frac{L_{Edd,s}}{R_s(\tau_{Edd})}
\end{equation}
which, in view of (13) has the solution
\begin{equation}
\Gamma_s(\tau) = \frac{1}{1 + z_s(\tau)} = (1-
\frac{2GM_s(\tau_{Edd})}{c^2R_s(\tau)_{Edd}})^{1/2} > 0
\end{equation}
which is consistent with Eq.s (9) and (11).
If one naively attributes Eddington limit luminosity to purely
thermal processes, one quickly finds that the required MECO surface
temperatures would be so high that photon energies would be far
beyond the pair production threshhold and the compactness would
assure that photon-photon collisions would produce numerous
electron-positron pairs. Thus the MECO surface region must be
dominated by a pair plasma. Pelletier \& Marcowith [1998] have shown
that the energy density of magnetic perturbations in equipartition
pair plasmas is preferentially converted to photon pressure, rather
than causing particle acceleration. The radiative power of an
equipartition pair plasma is proportional to $B^4$, (pair density
$\propto B^2$ and synchrotron energy production $\propto B^2$.)
Lacking the equipartition pair plasma, magnetic stress, $B^2/8\pi$,
and gravitational stress, $GM\rho/R$, on mass density $\rho$, would
both increase as $R^{-4}$ during gravitational collapse. Magnetic
fields much below equipartition levels would be incapable of
stopping the collapse. However, since photon pressure generated by
the pairs at equipartition increases more rapidly than gravitational
stresses, it is possible to stabilize the rate of collapse at an
Eddington limit rate. With this extremely efficient photon-photon
pair production mechanism, the radiation temperature and pressure is
buffered near the pair production threshold by two types of highly
redshifted quantum electrodynamic phase transitions which convert
photons into pairs on the MECO surface. The first one involves
optically thick photon-photon pair production while the second one
occurs for MECO surface magnetic fields strong enough to create additional
bound pairs in the surface plasma of the MECO. In the context
of an Eddington limited balance generated by the former process, the
latter process can lead to excess production of pairs, followed by
excess photon pressure and an expansion of the MECO surface. In this
manner the MECO Eddington limited collapse rate is inherently stable
(see Appendix 9 and 10). Stability is maintained by increased
(decreased) photon pressure ($\propto B^4$) if the field is
increased (decreased) by compression or expansion. For equipartition
conditions, the field also exceeds that required to confine the pair
plasma. Since the photon luminosity is not confined to the core it
will not be trapped, as occurs with neutrinos, however, the
radiation should be thermalized as it diffuses through an optically
thick environment. \textit{To reduce the field to the distantly
observed levels implied by our analysis of GBHC and AGN observations
would require the existence of a red shift of $z = 1.5\times
10^8(M/7M_\odot)^{1/2}$ (see Appendixes 9 and 10)}. The residual,
distantly observable magnetic moment and extremely faint, redshifted
radiations would be the only things that would distinguish such an
object from a black hole. \footnote{An additional point of support
for very large values of redshift concerns neutrino transport in
stellar core collapse. If a diffusion limited neutrino luminosity of
$\sim 10^{52}$ erg/s [Shapiro \& Teukolsky 1983] were capable of
very briefly sustaining a neutrino Eddington limit rate of collapse,
then the subsequent reduction of neutrino luminosity as neutrino
emissions are depleted and trapped in the core would lead to an
adiabatic collapse, magnetic flux compression, and photon emissions
reaching an Eddington limit. At this point the photon luminosity
would need to support a smaller diameter and more tightly
gravitationally bound mass. A new photon Eddington balance would
thus require an escaping luminosity reduced by at least the $\sim
10^{20}$ opacity ratio $(\sigma_T/\sigma_\nu)$, where $\sigma_T =
6.6\times 10^{-25}$ cm$^2$ is the Thompson cross section and
$\sigma_\nu = 4.4\times 10^{-45}$ cm$^2$ is the neutrino scattering
cross-section. Thus $L_\infty <10^{31-32}$ erg/s would be required.
For this to correspond to an Eddington limit luminosity as distantly
observed would require $1+z \sim 10^8$. The adiabatic relaxation of
neutrino support and formation of a pair plasma is an important step
in gravitational collapse that is not encompassed by polytropic
equation of state models of collapse. It is of some interest that if
neutrinos have non-zero rest mass they
might be trapped inside the photon sphere anyway.}

An electron-positron pair atmosphere of a MECO is an extremely
significant structure that conveys radiation from the MECO surface
to a zone with a much lower red shift and larger escape cone from
which it escapes. In order to describe this process
computationally within a numerical grid, a radial grid interval no
larger than $\sim 10^{-8}R_g$ would be needed, where $R_g =
GM/c^2$ is the gravitational radius. Although there have been many
numerical studies of the behavior of collapsing compact objects in
GR, to our knowledge none have sufficient numerical resolution to
examine the extreme red shift regime associated with MECO nor have
they considered the emergent properties of equipartition magnetic
fields and pair plasmas at high red shift. Until computer models
of gravitational collapse encompass these crucial physical and
computational elements, simulations that apparently produce black
hole states must be regarded as mere
scientific nonsense.

\section{Appendix 5 - The Quiescent MECO}

The quiescent luminosity of a MECO originates deep within its photon
sphere. When distantly observed it is diminished by both
gravitational red shift and a narrow exit cone. The gravitational
red shift reduces the surface luminosity by $1/(1+z)^2$ while the
exit cone further reduces the luminosity by the factor $27
R_g^2/(R(1+z))^2 \sim 27/(4(1+z)^2)$ for large z. Here we have used
\begin{equation}
\frac{R_g}{R} = \frac{1}{2}(1 - \frac{1}{(1+z)^2})~< \frac{1}{2}
\end{equation}
where R and z refer to the location from which photons escape.
The net outflow fraction of the luminosity provides the support for
the collapsing matter, thereby dynamically maintaining the EEP
requirement
of timelike world line completeness. The photons which finally escape do
so
from the photosphere of the pair atmosphere. The fraction of
luminosity from the MECO surface
that escapes to infinity in Eddington balance is
\begin{equation}
(L_{Edd})_s = \frac {4\pi G M_s c(1+z)}{\kappa} = 1.27 \times
10^{38}m(1+z_s)~~~~ erg/s
\end{equation}
where $m=M/M_\odot$. The distantly observed luminosity is:
\begin{equation}
L_\infty = \frac{(L_{Edd})_s}{(1+z_s)^2} = \frac {4\pi G M_s
c}{\kappa(1+z_s)}
\end{equation}
When radiation reaches the photosphere, where the temperature is $T_p$,
the fraction that escapes to be distantly observed is:
\begin{equation}
L_\infty = \frac{4 \pi R_g^2 \sigma T_p^4 27}{(1+z_p)^4}
= 1.56\times 10^7 m^2 T_p^4 \frac{27}{(1+z_p)^4}~~~~erg/s
\end{equation}
where $\sigma = 5.67\times 10^{-5}$ erg/s/cm$^2$ and subscript p
refers to conditions at the photosphere. The above equations yield:
\begin{equation}
T_\infty = T_p/(1+z_p) = \frac{2.3\times 10^7}{(m(1+z_s))^{1/4}}~~~~
K.
\end{equation}
To examine typical cases, a $10 M_\odot$, $m = 10$ GBHC modeled in
terms of a MECO with $z = 1.5\times 10^8(m/7)^{1/2}$ would have
$T_\infty = 1.1\times 10^5 K = 0.01$ keV, a bolometric luminosity,
excluding spin-down contributions, of $L_\infty =7.3\times 10^{30}
erg/s$, and a spectral peak at 220 A$^0$, in the photoelectrically
absorbed deep UV. For an m=$10^7$ AGN, $T_\infty = 630 K$,
$L_\infty = 7.2\times 10^{33} erg/s$ and a spectral peak in the
infrared at 4$\mu$. (Sgr A$^*$ at $m \approx 3\times 10^6$, would
have $T_\infty =1100$ K, and a 2.2 micron brightness below $0.6$
mJy; more than an order of magnitude below the observational upper
limit of 9 mJy [Reid et al. 2003].) Hence passive MECO without
active accretion disks, although not black holes, have lifetimes
much greater than a Hubble time and emit highly red shifted
quiescent thermal spectra that may be quite difficult to observe.
There are additional power law components of similar magnitude
that originate as magnetic dipole spin-down radiation.
Escaping radiation passes through a pair plasma atmosphere that can
be shown, \textit{ex post facto} to be radiation dominated
throughout. Under these circumstances, the radiation pressure within
the equilibrium atmosphere obeys $P_{rad}/(1+z) = constant$.
\footnote{Due to its negligible mass,we consider the pair atmosphere
to exist external to the MECO. Due to the slow collapse, the
exterior Vaidya metric can be approximated by exterior, outgoing
Finkelstein coordinates. In this case, the hydrostatic balance
equation within the MECO atmosphere is $\frac{\partial p}{\partial
r} = -\frac{\partial \ln{(g_{00})}}{2 \partial r}(p + \rho c^2)$,
where $g_{00} = (1-2R_g/r)$ and $\rho c^2 << p$. This integrates to
$p/(1+z) = constant$.} Thus the relation between surface and
photosphere temperatures is $T_s^4/(1+z_s) = T_p^4/(1+z_p)$. At the
MECO surface, we expect a pair plasma temperature of $T_s \approx
m_ec^2/k \sim 6\times 10^9 K$ because an equipartition magnetic
field effectively acts as a thermostat which buffers the temperature
of the optically thick synchrotron radiation escaping from the MECO
surface [Pelletier \& Marcowith 1998]. But since $T_\infty =
T_p/(1+z_p)$, and using $T_s=6\times 10^9$ K, we have that
\begin{equation}
T_p = T_s(\frac{T_s}{T_\infty (1+z_s)})^{1/3} = 3.8\times
10^{10}\frac{m^{1/12}}{(1+z_s)^{1/4}}=4.5\times 10^8
(m/7)^{-1/24}~K
\end{equation}
In the last expression, we have used $1+z_s=1.5\times
10^8(m/7)^{1/2}$. Using the above equations, this leads immediately
to $(1+z_p)=3500\times (m/7)^{1/3}$, independent of the surface
redshift, thus confirming that for MECO with pair atmospheres to
exist, they must be inherently highly redshifted. Due to the very
weak dependence of $T_p$ on $m$, the photosphere temperatures of
MECO are all very nearly $4.5\times 10^8$ K.
\section{Appendix 6 - An Actively Accreting MECO}
>From the viewpoint of a distant observer, accretion would
deliver mass-energy to the MECO, which would then radiate most of it
away. The contribution from the central MECO alone would be
\begin{equation}
L_\infty = \frac {4\pi G M_s c}{\kappa(1+z_s)}+ \frac{\dot{m}_\infty
c^2}{1+z_s}(e(1+z_s)-1)
= 4 \pi R_g^2 \sigma T_p^4 \frac {27} {(1+z_p)^4}
\end{equation}
where $e = E/mc^2 = 0.943$ is the specific energy per particle
available after accretion disk flow to the marginally stable orbit radius,
$r_{ms}$. Assuming that $\dot{m}_\infty$ is some fraction, f, of the
Newtonian Eddington limit mass accretion rate, $4\pi G M c/\kappa$, then
\begin{equation}
1.27\times 10^{38}\frac{m\eta}{1+z_s} =
(27)(1.56\times 10^7)m^2(\frac{T_p}{1+z_p})^4
\end{equation}
where $\eta=1+f((1+z_s)e -1)$ includes both quiescent and accretion
contributions to the luminosity. Due to the extremely strong
dependence on temperature of the density of pairs, (see Appendix 9
and 10) it is unlikely that the temperature of the photosphere will
be greatly different from the average of $4.6\times 10^8 K$ found
previously for a typical GBHC. Assuming this to be the case, along
with $z=10^8$, $m=10$, and $f=1$, we find $T_\infty = T_p/(1+z_p) =
1.3\times 10^7 K$ and $(1+z_p) = 35$, which indicates considerable
photospheric expansion. The MECO luminosity would be approximately
Newtonian Eddington limit at $L_\infty = 1.2\times 10^{39}$ erg/s.
For comparison, the accretion disk outside the marginally stable
orbit at $r_{ms}$ (efficiency = 0.057) would produce only $6.8\times
10^{37}$ erg/s. Thus the high accretion state luminosity of a GBHC
would originate primarily from the central MECO. The thermal
component would be `ultrasoft' with a temperature of only $1.3\times
10^7 K$ (1.1 keV). A substantial fraction of the softer thermal
luminosity would be Compton scattered to higher energy in the
plunging flow inside $r_{ms}$. Even if a disk flow could be
maintained all the way to the MECO surface, where a hot equatorial
band might result, the escaping radiation would be spread over the
larger area of the photosphere due
to photons origins deep inside the photon orbit.

For radiation passing through the photosphere
most photons would depart with some azimuthal momentum on spiral
trajectories
that would eventually take them across and through the accretion disk.
Thus a very large fraction of the soft photons would be subject
to bulk comptonization in the plunging region inside $r_{ms}$.
This contrasts sharply with the situation for neutron stars
where there probably is no comparable plunging region and
few photons from the surface cross the disk. This
could account for the fact that hard x-ray spectral tails are
comparatively much stronger for high state GBHC.
Our preliminary calculations for photon trajectories
randomly directed upon leaving the
photon sphere indicate that this process
would produce a power law component with photon index greater than 2.
These are difficult, but important calculations for which the effects
of multiple scattering must be considered. But they are beyond the
scope of this paper, which is intended as a first description of the
general MECO model.

\section{Appendix 7 - Magnetosphere - Disk Interaction}

In LMXB, when the inner disk engages the magnetosphere, the inner
disk temperature is generally high enough to produce a very
diamagnetic plasma. This may not be the case for AGN. Surface
currents on the inner disk distort the magnetopause and they also
substantially shield the outer disk such that the region of strong
disk-magnetosphere interaction is mostly confined to a ring or
torus, of width $\delta r$ and half height $H$. This shielding
leaves most of the disk under the influence of its own internal
shear dynamo fields, [e.g. Balbus \& Hawley 1998, Balbus 2003]. At
the inner disk radius the magnetic field of the central MECO is
much stronger than the shear dynamo field generated within the
inner accretion disk. In MHD approximation, the force density on
the inner ring is $F_v = (\nabla \times B) \times B / 4\pi$. For
simplicity, we assume coincident magnetic and spin axes of the
central object and take this axis as the $z$ axis of cylindrical
coordinates $(r,\phi,z)$.
The magnetic torque per unit volume of plasma in the inner ring of
the disk that is threaded by the intrinsic magnetic field of the
central object, can be approximated by $\tau_v=rF_{v\phi} = r
\frac{B_z}{4\pi} \frac{\partial B_{\phi}}{\partial z} \sim r
\frac{B_zB_{\phi}}{4\pi H}$, where $B_{\phi}$ is the average
azimuthal magnetic field component. We stress that $B_{\phi}$, as
used here, is an average toroidal magnetic field component. The
toroidal component likely varies episodically between reconnection
events [Goodson \& Winglee 1999, Matt et al. 2002, Kato, Hayashi
\& Matsumato 2004, Uzdensky 2002].
The average flow of disk angular momentum entering the inner ring
is $\dot{M}r v_k$, where $\dot{M}$ is mass accretion rate and
$v_k$ is the Keplerian speed in the disk. This angular momentum
must be extracted by the magnetic torque, $\tau$, hence:
\begin{equation}
\tau = \dot{M}r v_k = r\frac{B_zB_{\phi}}{4\pi H}(4\pi r H \delta
r).
\end{equation}
In order to proceed further, we assume that $B_{\phi} = \lambda
B_z$, $B_z=\mu/r^3$, and use $v_k=\sqrt{GM/r}$, where $\lambda$ is
a constant, presumed to be of order unity, $\mu$ is the magnetic
dipole moment of the central object $M$, its mass, and $G$, the
Newtonian gravitational force constant. With these assumptions we
obtain
\begin{equation}
\dot{M} = (\frac{\lambda \delta r}{r}) \frac {\mu^2}{r^5 \omega_k}
\end{equation}
where $\omega_k = v_k/r$ and the magnetopause radius, $r_m$ is
given by
\begin{equation}
r_m = (\frac{\lambda \delta
r}{r})^{2/7}(\frac{\mu^4}{GM\dot{M}^2})^{1/7}
\end{equation}

In order to estimate the size of the boundary region, $(\delta r
/r)$, we normalized this disk-magnetosphere model for agreement
with radii calculated for an elaborate model of a gas pressure
dominated disk [Ghosh \& Lamb 1992]. Although we find the portion
of the inner disk threaded by magnetic fields to be smaller than
the Ghosh \& Lamb model, this size for the inner radius yields
very accurate results for accreting millisecond pulsars, which
have known magnetic moments. We find $(\frac{\lambda \delta r}{r})
= 0.015$. Using units of $10^{27}$ gauss cm$^3$ for magnetic
moments, $100$ Hz for spin, $10^6$ cm for radii, $10^{15}$ g/s for
accretion rates, solar mass units, $\lambda \delta r/r = 0.015$
and otherwise obvious notation, we find the magnetosphere radius
to be:
\begin{equation}
r_m~=~4\times 10^6 {(\frac{\mu_{27}^4}{m
\dot{m}_{15}^2})}^{1/7}~~~~
cm
\end{equation}
where $m=M/M_\odot$ and the disk luminosity is
\begin{equation}
L = \frac{GM\dot{m}}{2r_m}
\end{equation}
The co-rotation radius, at which disk Keplerian and magnetosphere
spins match is:
\begin{equation}
r_c = 7\times 10^6{(\frac{m}{\nu_2^2})^{1/3}} ~~~cm
\end{equation}
The low state disk luminosity at the co-rotation radius is the
maximum luminosity of the true low state and is given by:
\begin{equation}
L_c = \frac{GM\dot{m}}{2r_c}=1.5 \times 10^{34} \mu_{27}^2
{\nu_2}^3 m^{-1}~~~~erg/s
\end{equation}
The minimum high state luminosity for all accreting matter being
able to reach the central object occurs at approximately the same
accretion rate as for $L_c$ and is given by:
\begin{equation}
L_{min} = \xi \dot{m}c^2 = 1.4\times 10^{36} \xi \mu_{27}^2
\nu_2^{7/3} m^{-5/3}~~~erg/s
\end{equation}
Where $\xi \sim 0.42$ for MECO for the photon sphere\footnote{The
time for a luminosity variation to be observed is very long for
energy released by processes inside the photon sphere.} and $\xi =
0.14$ for NS is the efficiency of accretion to the central surface.

For a MECO in true quiescence, the inner disk radius is larger than
the light cylinder radius. In NS, and GBHC and AGN modeled as MECO,
the inner disk may be ablated due to radiation from the central
object. The inner disk radius can be ablated to distances larger
than $5\times 10^4 km$ because optically thick material can be
heated to $\sim 5000 K$ and ionized by the radiation. The maximum
disk luminosity of the true quiescent state occurs with the inner
disk radius at the light cylinder, $r_{lc}=c/\omega_s= r_m$. The
maximum luminosity of the quiescent state is typically a factor of a
few larger than the average observed quiescent luminosity.
\begin{equation}
L_{q,max} = (2.7\times
10^{30} erg/s) \mu_{27}^2 \nu_2^{9/2} m^{1/2}
\end{equation}

We calculate the average quiescent luminosities in the soft x-ray
band from $\sim 0.5 - 10$ keV using the correlations of Possenti
et al. [2002] with spin-down energy loss rate as:
\begin{equation}
L_q = \beta \dot{E} = \beta 4 \pi^2 I \nu \dot{\nu}
\end{equation}
where $I$ is the moment of inertia of the star, $\nu$ its rate of
spin and $\beta$ a multiplier that can be determined from this new
$\dot{E} - L_q$ correlation for given $\dot{E}$; i.e., known spin
and magnetic moment. In previous work we had used $\beta = 10^{-3}$
for all objects, but $\beta \sim 3\times 10^{-4}$ would be the
average value, for GBHC and AGN modeled as MECO, consistent with the
Possenti correlation. We assume that the luminosity is that of a
spinning magnetic dipole for which $\dot{E} = 32\pi^4 \mu^2
\nu^4/3c^3$, (Bhattacharya \& Srinivasan 1995] where $\mu$ is the
magnetic moment. Thus the quiescent x-ray luminosity would then be
given by:
\begin{equation}
L_q = \beta \times \frac{32 \pi^4 \mu^2 \nu^4}{3c^3} = 3.8 \times
10^{33} \beta \mu_{27}^2 \nu_2^4 ~~~~~erg/s
\end{equation}
According to the Possenti correlation, $\beta = L_q/\dot{E}~~
\propto~ \dot{E}^{0.31}$. $\beta$ should be a dimensionless,
ratio, and independent of mass. But since $\dot{E}$ is
proportional to mass, we extend the Possenti relation, without
loss of generality, to provide a mass scale invariant quantity. We
therefore take $\beta \propto (\dot{E}/m)^{0.31}$. From the
Possenti correlation, assuming all the objects in their study have
the canonical $m=1.4$, we then find that
\begin{equation}
\beta=7\times 10^{-4}(\dot{E}/m)^{0.31}=4.6\times
10^{-4}(10^{-36}L_c\nu_2)^{0.31}
\end{equation}

Since the magnetic moment, $\mu_{27}$, enters each of the above
luminosity equations it can be eliminated from ratios of these
luminosities, leaving relations involving only masses and spins. For
known masses, the ratios then yield the spins. Alternatively, if the
spin is known from burst oscillations, pulses or spectral fit
determinations of $r_c$, one only needs one measured luminosity,
$L_c$ or $L_{min}$ at the end of the transition into the soft state,
to enable calculation of the remaining $\mu_{27}$ and $L_q$. In the
case of GBHC modeled as MECO, we found it to be necessary to
estimate the co-rotation radius from multicolor disk fits to the
thermal component of low state spectra. The reason for this is that
the luminosities are sometimes unavailable across the whole spectral
hardening transition from $L_c$ to $L_{min}$ for GBHC. 
For GBHC, it is a common finding that the low state inner disk
radius is much larger than that of the marginally stable orbit;
e.g. [Markoff, Falcke \& Fender 2001, $\dot{Z}$ycki, Done \& Smith
1997a,b 1998, Done \& $\dot{Z}$ycki 1999, Wilson \& Done 2001].
The presence of a magnetosphere is an obvious explanation. Given
an inner disk radius at the spectral state transition, the GBHC
spin frequency follows from the Kepler
relation $2 \pi \nu_s = \sqrt{GM/r^3}$.
Although we have taken our model and used it to predict the spin
rates and accurate quiescent luminosities for NS and GBHC, it now
appears that we could use the fact that the model fits well to
calculate more accurate parameters. By placing the last mass scale
invariant Possenti relation for $\beta$ into the relation for
quiescent luminosity and using it with the expression for $L_c$, we
can determine spin rates to be given by
\begin{equation}
\nu = 89(L_{q,32}/(mL_{c,36}^{1.31}))^{1/1.31}~~~Hz
\end{equation}
where $L_{q,32} = 10^{-32}L_q$ and $L_{c,36} = 10^{-36}L_c$.
Consistency of the above mass scaling equations also requires that
the magnetic field equipartition condition must hold as
\begin{equation}
L_{c,36}/(m^{4}\nu^{3})=1
\end{equation}
Using the above equations the average spin rate for the GBHC is
found to be 10 Hz and the average GBHC magnetic moment has the
average value of 2200 gauss-cm$^3$. Since this scaling technique
essentially preserves the quiescent luminosities, the calculated
results are more reliable than those obtained from determining the
inner disk radii from spectral fitting.

\section{Appendix 8 - Properties of The Low Hard State - Mass Ejection
and
Radio
Emission}

The radio flux, $F_{\nu}$, of jet sources has a power law
dependence on frequency of the form
\begin{equation}
F_{\nu}~ \propto~ \nu^{-\alpha}
\end{equation}
It is believed to originate in jet outflows and has been shown to
be correlated with the low state x-ray luminosity [Merloni, Heinz
\& DiMatteo 2003], with $F_\nu \sim L_x^{0.7}$. The radio
luminosity of a jet is a function of the rate at which the
magnetosphere can do work on the inner ring of the disk. This
depends on the relative speed between the magnetosphere and the
inner disk; i.e., $\dot{E} =\tau (\omega_s - \omega_k)$, or
\begin{equation}
\dot{E} = (\frac{\lambda \delta r}{r})\frac{\mu^2 \omega_s (1 -
\frac{\omega_k}{\omega_s})}{r^3}
~\propto~ \mu^2 M^{-3}\dot{m}_{Edd}^{6/7}\omega_s(1 -
\frac{\omega_k}{\omega_s})
\end{equation}
Here $\dot{m}_{Edd}$ is the mass accretion rate divided by the
rate that
would produce luminosity at the Eddington limit for mass $M$.
Disk mass, spiraling in quasi-Keplerian orbits from negligible
speed at radial infinity must regain at least as much energy as
was radiated away in order to escape. For this to be provided by
the magnetosphere requires $\dot{E} \geq GM\dot{M}/2r$, from which
$\omega_k \leq 2\omega_s/3$. Thus the magnetosphere alone is
incapable of completely ejecting all of the accreting matter once
the inner disk reaches this limit and the radio luminosity will be
commensurately reduced and ultimately cut off at maximum x-ray
luminosity for the low state and $\omega_k=\omega_s$. Typical data
for GX339-4 [Gallo, Fender \& Pooley 2003] are shown in Figure 1.
For very rapid inner disk transit through the co-rotation radius,
fast relative motion between inner disk and magnetosphere can heat
the inner disk plasma and strong bursts of radiation pressure from
the central object may help to drive large outflows while an
extended jet structure is still largely intact. This process has
been calculated \footnote{though for inner disk radii inside the
marginally stable orbit [Chou \& Tajima 1999]}using pressures and
poloidal magnetic fields of unspecified origins. A MECO is
obviously capable of supplying both the field and a radiation
pressure. The hysteresis of the low/high and high/low state
transitions may be associated with the need for the inner disk to
be completely beyond the corotation radius before a jet can be
regenerated after it has subsided.
Since $\dot{E} \propto r^{-3}$ and $L_d \propto r^{-9/2}$, it is
apparent that we should expect radio luminosity, $L_R \propto
L_d^{2/3}$. In particular we find
\begin{equation}
L_R = C(M,\beta, \omega_s)2L_c^{1/3}
L_d^{2/3}(1-\omega_k/\omega_s)
\end{equation}
where $\beta =\mu/M^3$ and $C(M,\beta, \omega_s)$ is a constant,
dependent on the radio bandpass. It has been analyzed and evaluated
[Robertson \& Leiter 2004]. The cutoff at $\omega_k=\omega$ is shown
by the line on Figure 1. The cutoff typically occurs with x-ray
luminosity of $\sim 0.01 - 0.02$ times Eddington luminosity. If we
let $x=L_d/L_c$, then for $x < 1$, corresponding to the low state,
then the above equations imply that:
\begin{equation}
L_R =C(M,\beta, \omega_s) 2L_c(x^{2/3} - x)
\end{equation}
The function has a maximum value of $0.3C(M,\beta, \omega_s)
L_c$ at $x=0.3$.
Strictly speaking, $L_d$, in these equations should be the
bolometric luminosity of the disk, however, the x-ray luminosity
over a large energy band is a very substantial fraction of the disk
luminosity. To compare with the correlation exponent of 2/3 obtained
here, recent studies, including noisy data for both GBHC and AGN
have yielded $0.71 \pm 0.01$ [Gallo, Fender \& Pooley 2003], 0.72
[Markoff et al. 2003], $0.60 \pm 0.11$ [Merloni, Heinz \& Di Matteo 2003]
and $0.64 \pm 0.09$[Maccarone, Gallo \& Fender 2003]. For $\alpha$ in
the range (0 to -0.5), $\beta \propto M^{-1/2}$, $\omega_s \propto M^{-1}$
and $L_c \propto M$, the MECO model yields $C(M) \propto M^{(9-4\alpha)/12}$
and (neglecting the cutoff region)
\begin{equation}
log L_R = (2/3)log L_x + (0.75 - 0.92) log M + const.
\end{equation} which is a better fit to the "fundamental plane" of
Merloni, Heinz \& Di Matteo [2003] than any of the ADAF,
disk/corona or disk/jet models they considered (see their
Figure 5 for a $\chi^2$ density plot). This last relation correctly
describes the correlation for both GBHC and AGN.
In the low state, the inner disk radius is inside the light
cylinder, with hot, diamagnetic plasma reshaping the magnetopause
topology [Arons et al. 1984]. This magnetic propeller regime
(Ilarianov \& Sunyaev 1975, Stella, White \& Rosner 1986, Cui 1997,
Zhang, Yu \& Zhang 1997, Campana et al. 1998] exists until the inner
disk pushes inside the co-rotation radius, $r_c$. From $r_{lc}$ to
$r_c$, the x-ray luminosity may increase by a factor of $\sim 10^3 -
10^6$. Inside $r_c$, large fractions of the accreting plasma can
continue on to the central object and produce a spectral state
switch to softer emissions. We have shown (Robertson \& Leiter 2002)
that magnetic moments and spin rates can be determined from
luminosities at the end points of the transition from low/hard to
high/soft spectral states. In that paper the magnetic moments and
spins were used to calculate the $\sim 10^{3-6}$ times fainter
quiescent luminosities expected from spin-down. During waning phases
of nova outbursts, $L_c \sim 0.02L_{Edd}$ can be identified as the
maximum disk luminosity upon entering the low state.

Until the inner disk reaches $r_c$, accreting plasma is ejected. It
may depart in a jet, or as an outflow back over the disk as plasma
is accelerated on outwardly curved or open magnetic field lines.
Radio images of both flows have been seen [Paragi et al. 2002].
Equatorial outflows could contribute to the low state hard spectrum
by bulk Comptonization of soft photons in the outflow, however, we
think the hard spectrum originates primarily in patchy coronal
flares [Merloni \& Fabian 2002] on a conventional geometrically
thin, optically thick disk consistent with the existence of a
magnetic flux line breaking and reconnection process acting above
the accretion disk.

\section{Appendix 9 - The Existence and Stability of Highly Redshifted
MECO } 

We have shown that the existence of AGN containing observable
intrinsic magnetic moments is consistent with a new class of
magnetospheric eternally collapsing object (MECO) solutions of the
Einstein field equations of General Relativity. These solutions are
based on a strict adherence to the SPE requirement for timelike
world line completeness; i.e., that the world lines of physical
matter under the influence of gravitational and non-gravitational
forces must remain timelike in all regions of spacetime. Since there
is nothing in the structure of the Einstein tensor, $G^{\mu \nu}$,
on the left hand side of the Einstein field equation that
dynamically enforces `time like world line completeness', we have
argued that the SPE constrains the physically acceptable choices of
the energy momentum tensor, $T^{\mu \nu}$ to contain
non-gravitational forces that can dynamically enforce it. In this
context we have found the long-lived MECO solutions.

An enormous body of physics scholarship developed primarily over the
last half century has been built on the assumption that trapped
surfaces leading to event horizons and curvature singularities
exist. Misner, Thorne \& Wheeler [1973], for example in Sec. 34.6
clearly state that this is an assumption and that it underlies the
well-known singularity theorems of Hawking and Penrose. In contrast,
we have found that strict adherence to the SPE demand for timelike
world line completeness requires a \textit{`no trapped surface
condition'}. This has led to the quasi-stable, high red shift MECO
solutions of the Einstein field equations. The physical mechanism of
their stable rate collapse is an Eddington balance maintained by the
distributed photon generation of a a highly compact and redshifted
equipartition magnetic field. This field also serves to confine the
pair plasma dominated outer layers of the MECO and the thin MECO
pair atmosphere. Red shifts of $z \sim 10^8(m/7)^{1/2}$ have been
found to be necessary for compatibility with our previously found
magnetic moments for GBHC.

In this Appendix we give a detailed description of the MECO's
properties. These properties can be shown to be compatible with
standard gas pressure dominated `alpha' accretion disks which is
consistent with the fact (previously shown in Appendix 7 and 8) that
the magnetosphere/disk interaction affects nearly all of the
spectral characteristics of NS and GBHC in LMXB systems and accounts
for them in a unified and complete way, including jet formation and
radio emissions. This model is solidly consistent with accreting NS
systems, for which intrinsic magnetic moments obtained from
spin-down measurements allow little choice. The magnetic fields are
too strong to ignore. Since the similar characteristics of GBHC are
cleanly explained by the same model, the MECO offers a unified
theory of LMXB phenomenology as well as extensions to AGN. Since
MECO lifetimes are orders of magnitude greater than a Hubble time,
they provide an elegant and unified framework for understanding the
broad range of observations associated with GBHC and AGN. Lastly we
have indicated some ways in which the existence of MECO in GBHC and
AGN might be detected and confirmed.

A MECO is, in many ways, more exotic than a black hole with its mere
mass and spin. It is equally compact but its surface magnetic field
is sufficient to produce pairs from the quantum electrodynamic
vacuum. This occurs at a threshold that is insensitive to mass, thus
MECO can range in mass from galactic black hole candidates GBHC to
active galactic nuclei AGN. The scaling with mass of the distantly
observed magnetic fields, $B \propto M^{-1/2}$ allows the ratio
$L_c/L_{Edd}$ to also be mass scale invariant [Robertson \& Leiter
2004]. The MECO interior magnetic fields are relatively modest.
Interior and surface fields differ due to substantial pair drift
currents on the MECO surface. In general, plasmas in hydrostatic
equilibrium in magnetic and gravitation fields experience drift
currents proportional to ${\bf g\times B}/B^2$. The general
relativistic generalization of this provides the key to our
understanding of the high redshifts of MECO.

The surface temperature and high luminosity to radius ratio
(hereafter L/R compactness, (see Appendix 10). of the MECO Eddington
limited, timelike, secular collapsing state implies that the plasma
is dominated by electron-positron pairs. These are generated by
colliding photons due to the optically thick synchrotron luminosity
of the intrinsic MECO magnetic field, both within the interior and
on the MECO surface. Recall that the surface is well inside the
photon orbit and the bulk of the photon outflow from the surface
falls back. The existence of the MECO state requires that:
\begin{equation}
L_{Edd}(outflow)\sim L_{Syn}(out)
\end{equation}
within the MECO and
\begin{equation}
L_{Edd,S}(escape) \sim L_{Syn,S}(escape)
\end{equation}
at the MECO surface S. Where (see Appendix 4)
\begin{equation}
L_{Edd,S}(escape) \sim (4\pi G M_s(\tau)c /\kappa)(1 + z_s) \sim
1.3\times10^{38}m(1+z_s)
\end{equation}
\begin{equation}
L_{Syn,S}(escape) \sim L_{Syn,S}(out)/{(4/27)(1+z_s)^2}
\end{equation}
Assuming a temperature near the pair production threshold, the
rate of synchrotron photon energy generation in a plasma
containing $N_\pm$ electrons and positrons is [Shapiro \&
Teukolsky 1983]
\begin{equation} L_{Sync,S}(out) \sim
(16e^4/3m_e^2c^3)N_\pm B^2(T_9/6)^2 \sim 1.27 \times 10^{-14}N_\pm
B^2(T_9/6)^2~~erg/s
\end{equation}
where $T_9=10^{-9}T$ and $T_9/6=kT/m_ec^2$.
For an Eddington equilibrium, we require the synchroton generation
rate to produce the outflow through the MECO surface. Thus
\begin{equation}
L_{Edd}(out) \sim 1.27 \times 10^{38} m (4(1+z_s)^3/27) \sim 1.27
\times 10^{-14}N_\pm B^2(T_9/6)^2~~erg/s
\end{equation}
which implies that
\begin{equation}
N_\pm B^2 \sim 10^{52}(m/7)(1 + z_s)^3(6/T_9)^2~~erg/cm^3
\end{equation}
\bigskip
>From Section 6 of Baumgarte \& Shapiro [2003], we note that if
$\mu$ is the distantly observed MECO magnetic moment and
$(1+z_s)>>1$ is the MECO surface redshift, then the Einstein-
Maxwell equations imply that the components of the MECO dipole
magnetic field strength, B at distance r are given by
\begin{equation}
B_r=2F(x)\mu \cos(\theta)/r^3
\end{equation}
and
\begin{equation}
B_{\theta}=G(x)\mu \sin(\theta)/r^3
\end{equation}
where $x=r_s/2R_g$ and
\begin{equation}
F(x)=(-3x^3)(ln(1-x^{-1})+x^{-1}(1+x^{-1}/2))
\end{equation}
\begin{equation}
G(x)=(6x^3)((1-x^{-1})^{1/2}ln(1-x^{-1})+x^{-1}(1-x^{-1}/2)(1-x^{-
1})^{-1/2}
)
\end{equation}
Note that for $r_s>>2R_g$, $x>>1$ and both $F(x)$ and $G(x)
\rightarrow 1$, while as we approach a compact MECO surface where
$(1+z_s) >>> 1$, then $x \rightarrow 1^+$ and
\begin{equation}
F(x) \rightarrow -3ln(1-x^{-1}) = -3ln(1/(1+z_s)^2) = 6ln(1+z_s)
\end{equation}
and
\begin{equation}
G(x)=3(1-x^{-1})^{-1/2}=3(1+z_s)
\end{equation}
Hence the radial component of the magnetic field on the MECO
surface is given by
\begin{equation}
B_{r,S^+} =12ln(1+z_s)\mu \cos(\theta)/(2R_g)^3
\end{equation}
while the poloidal component is given by
\begin{equation}
B_{\theta,S^+}=3\mu(1+z_s)\sin(\theta)/(2R_g)^3
\end{equation}
The interior magnetic dipole fields, $B'$ in the MECO, which are
due to the interior MECO magnetic dipole moment $\mu(r)$ in the
interior will be given by
\begin{equation}
B'_r=12\mu(r) \cos(\theta)\ln(1+z)/r^3
\end{equation}
and
\begin{equation}
B'_{\theta}=3 \mu(r) \ln(1+z) \sin(\theta)/r^3
\end{equation}
Thus the expressions for the exterior magnetic field just outside of
the MECO surface differ by the factors $F(x_s)$ and $G(x_s)$ from
those of the interior magnetic field. But these general relativistic
expressions imply important consequences because:
(a) The general relativistic structure of the Maxwell-Einstein
equations causes the radial and poloidal exterior components of the
MECO magnetic dipole fields to undergo redshift effects which are
different functions of $(1 + z_s)$\\ and
(b) The radial component $B_{r,S}$ of the magnetic dipole field is
continuous at the MECO surface, but the poloidal component is not.
$B_{\theta,S^+}$ is different from $B'_{\theta,S^-}$ at the MECO
surface.

This difference is caused by powerful $e^{\pm}$ drift currents that
are induced by the strong gravitational field and enhanced by the
differing general relativistic dependence on redshift of the
poloidal and radial magnetic field components. This is a general
relativistic generalization of the fact that a plasma in hydrostatic
equilibrium in gravitational and magnetic fields experiences drift
currents proportional to ${\bf g\times B}/B^2$. In fact, it is the
drift currents that generate the distantly observed magnetic moments
seen in MECO-GBHC and MECO-AGN.

The MECO magnetic moment coupling to a surrounding accretion disk
will cause it to be a slow rotator. Hence to estimate the strength
of the magnetic field just under the surface of the MECO of the GBHC
we can use the results obtained for low redshift, slowly rotating
compact stellar objects with magnetic fields. The magnetic field
strength in the interior of a slowly rotating neutron star of radius
$\sim 10$ km, was shown to be $\sim 10^{13}$G [Gupta, Mishra, Mishra
\& Prasanna 1998]. When scaled to the $\sim 7 M_\odot$ and $R \sim
2R_g$ size of the MECO, the magnetic fields under the surface can be
estimated to be (neglecting latitude angle dependence):
\begin{equation}
B_{r,S^-} \sim (2\mu/(2R_g)^3)6 ln(1+z_s) \sim (10^{13.7} gauss)/
(M / 7M_\odot)^{1/2}
\end{equation}
\begin{equation}
B_{\theta,S^-} \sim (\mu/(2R_g)^3)6 ln (1+z_s) \sim (10^{13.4}
gauss)/ (M / 7M_\odot)^{1/2}
\end{equation}
Then from Eq.s (66)-(73), the exterior magnetic fields on the MECO
surface S are:
\begin{equation}
B_{r,S^+} = B_{r,S^-} \sim (2\mu/(2R_g)^3)6 ln(1+z_s)cos(\theta)
\sim (10^{13.7} gauss) cos(\theta) /(M/7M_\odot)^{1/2}
\end{equation}
\begin{equation}
B_{\theta,S^+} \sim (\mu / (2R_g)^3)3 (1+z_s) sin(\theta)
\end{equation}
Using these equations, the magnitude of the surface redshift
$(1+z_s)$ for a MECO can be directly determined by noting that the
strength of the poloidal component of the equipartition magnetic
field $B_{\theta,S}$ on the MECO surface must be the same for all
MECO's (i.e. it must be mass scale invariant since it cannot be
locally much larger than the quantum electrodynamically determined
maximum value for a NS given by $B_{\theta,S} \sim 10^{20}$ gauss
[Harding, A., 2003]). This is because surface magnetic fields much
larger than $\sim 10^{20}$ gauss would create a spontaneous quantum
electrodynamic phase transition associated with the vacuum
production of bound pairs on the MECO surface [Zaumen 1976]. This
would cause more pairs to be produced than those required by the
Eddington balance of the MECO surface. This would then cause the
MECO surface to expand. However the resultant expansion due to this
process would reduce the redshift and the surface poloidal magnetic
field thus quenching the vacuum production of bound pairs and
allowing the MECO surface to contract. This stability mechanism on
the MECO surface implies that its surface redshift $(1+z_s)$ can be
dynamically determined from the preceding pair of equations.
Neglecting the trigonometric functions common to both sides of the
equations, the ratio of these external field components in Eqs. (69)
and (71), yields
\begin{equation}
3(1+z_s)/[6ln(1+z_s)] \sim 10^{20}/[10^{13.4}/(m/ 7)^{1/2}]
\end{equation}
for which the solution is
\begin{equation}
(1+z_s) \sim 1.5\times 10^8(m/7)^{1/2}
\end{equation}
In addition we obtain
\begin{equation}
\mu / (2R_g)^3 \sim (2.2\times 10^{11} gauss)/(m / 7)^{1/2}
\end{equation}
{which implies that the average distantly observed intrinsic
magnetic moment of the MECO is
\begin{equation}
\mu \sim (2\times 10^{30}gauss~ cm^3)(m / 7)^{5/2}
\end{equation}
This is in good agreement with our analysis of observations.
>From the above equations we can now get a rough estimate of the pair
density, by considering the $N_\pm$ to be uniformly distributed over
a volume of $4(1+z_s) \pi (2R_g)^3/3$ and by considering the
interior magnetic field to be uniform at $2.5\times
10^{13}/(m/7)^{1/2}~gauss$. Hence
\begin{equation}
n_\pm=10^{22} (m/7)^{-3/2}(6/T_9)^2~~/cm^3
\end{equation}
\bigskip
which for a GBHC, agrees within a factor of two of the result
found for a pair plasma at $T=6\times 10^9$K. This result for
$n_\pm$ implies that surface temperature would increase with
increasing mass, however, it only increases by a factor of $10$
for $m=10^8$. Since mean MECO
densities scale as $1/m^2$, one might expect larger density
gradients and different ratios of pairs to neutrons and protons in
AGN compared to GBHC, which are approximately
of nuclear densities.

Azimuthal MECO surface currents are the source of the distantly
observed magnetic moment seen in the MECO-GBHC. The magnitude of
these surface currents is essentially mass scale invariant and is
given by
\begin{equation}
i(S) = (c / 4\pi ) (\mu /(2R_g)^3)[3(1+z_s)]sin(\theta) \sim
3.3\times 10^{26}~~amp
\end{equation}
total current on GBHC surface, which corresponds to
\begin{equation} \sim 2\times 10^{45} e^\pm /sec
\end{equation}
combined surface e(+ -) flow. Hence the above equations imply that
the corresponding drift speeds of electrons and positrons are $v/c
\sim 1$. This implies that the opposed e(+-) pairs currents on the
MECO surface are moving relativistically and hence will have a very
long lifetime before annihilating. This maintains a stable flow of
current as required to generate the distantly observed
MECO magnetic moments.
The radiation pressure at the outer surface of the MECO is
\begin{equation}
P_{Syn}(out)= L_{Syn}(out)/[4\pi(2R_g)^2c]=
\sim 2.7\times 10^{38}(m/7)^{1/2}~~erg/cm^3
\end{equation}
For comparison, the mass-energy density of a MECO is
\begin{equation}
\rho c^2 \sim Mc^2/[(1+z_s)4\pi (2R_g)^3/3] \sim 2\times
10^{27}/(m/7)^{2.5}
\end{equation}
\bigskip
which suggests that MECO is radiation dominated and very tightly
gravitationally bound. An order of magnitude calculation of binding
energy \footnote{This has obvious, important consequences for
hypernova models of gamma-ray bursters.} yields $\sim 1.5 Mc^2
~ln(1+z_s)$ for a residual mass, M. Thus the progenitor of a
MECO-GBHC would have a mass of 200-300 $M_\odot$, and this suggests
that they would be among the earliest and most massive stars in the
galaxy. On the other hand the progenitor of a $10^9 M_\odot$
MECO-AGN with its very large intrinsic magnetic moment would most
likely originate from the collapse of magnetized plasma containing angular
momentum in addition to aggregates of primordial
planetoids, like that discussed in section 5 of this paper, with a
total mass of $4\times 10^{10}M_\odot$.\\
\bigskip

\section{Appendix 10 - The $\gamma + \gamma \leftrightarrow e^\pm
$Phase
Transition
and MECO Existence }

It is well-known that a spherical volume of radius $R$ containing
a luminosity $L_{\gamma}$ of gamma ray photons with energies $>$
1 MeV,\footnote{1 MeV photons correspond to $T \sim 10^{10}~K$,
which is only slightly beyond the pair threshold, and easily
within reach in gravitational collapse.} will become optically
thick to the $\gamma + \gamma \leftrightarrow e^\pm$ process when
\begin{equation}
\tau_\pm \sim n_\gamma\sigma_{\gamma\gamma}R \sim 1
\end{equation}
and
\begin{equation}
n_\gamma \sim L_\gamma/(2\pi R^2 m_ec^3)
\end{equation}
is the number density of $\gamma$-ray photons with energies $\sim$
1 MeV, $L_\gamma$ is the $gamma$-ray luminosity, $R$ is the radius
of the volume, and $\sigma{\gamma\gamma}$ is the pair production
cross section.
Since $\sigma{\gamma\gamma} \sim \sigma_T$ near threshold, it
follows that the system becomes optically thick to photon-photon
pair production when the numerical value of its compactness
parameter $L_\gamma/R$ is
\begin{equation}
L_\gamma/R \sim 4\pi m_e c^3/\sigma_T \sim 5\times 10^{29}~~~
erg/cm-sec
\end{equation}
Hence $\tau_\pm > 1$ will be satisfied for systems with
compactness
\begin{equation}
L_\gamma/R > 10^{30}~~erg/cm-sec
\end{equation}
For an Eddington limited MECO, which has a very large surface
redshift $(1 + z) >> 1$ at $R \sim 2R_g$ , and taking the
proper length and volume into consideration, the optical depth to
photon-photon pair production has the very large value
\begin{equation}
\tau_\pm(1+z) \sim (L_{\gamma, 30}/R) \sim 10^2 \times (1+z)^3
>> 1~~erg/cm-sec
\end{equation}
Thus the resultant $\gamma + \gamma \leftrightarrow e^\pm$ phase
transition in the MECO magnetic field, $B_S$, creates an optically
thick pair dominated plasma. Taking the photon escape cone factor
$\sim 1 /(1+z)^2$ into account, the process generates a net
outward non-polytropic radiation pressure
\begin{equation}
P \sim F(1 + z) B_S^4 m
\end{equation}
on the MECO surface. The increase of pressure with redshift is a key
feature of the Eddington limited secular balance at $R \sim
2GM/c^2$. Thus trapped surfaces, which lead to event horizons, can
be prevented from forming. As discussed in Appendix 9 above., the
balance is mass scale invariantly stabilized at the threshold of
magnetically produced pair breakdown of the vacuum.\\

\section{Appendix 11 - On The Black Hole Kerr-Schild Metric And
MECO Vaidya Metric Solutions To the Gravitational Collapse Problem}

In discussions with experts in general relativity the validity our
motivation to look for physical alternatives to black holes has been
questioned. Our work has been based on the assumption that
the preservation of the Strong Principle of Equivalence (SPOE) in
Nature implies that metrics with event horizons are non-physical.
The objection to this has been based on the well known fact that
for massive particles under the action of both gravitational and
non-gravitational forces, the timelike nature of the world line of
massive particles is preserved. The generally covariant equation of motion
for their timelike world lines in spacetime is given by
\begin{equation}
Du^{\mu} / d\tau = a^{\mu} = K^{\mu}
\end{equation}
Here $u^{\mu}$ is the four velocity of the massive particle and
$K^{\mu}$ is the generally covariant non-gravitational four-vector
force which in general relativity is required to obey the dynamic
condition
\begin{equation}
K^{\mu}a_{\mu} = 0
\end{equation}
Then from the above two equations it follows that
\begin{equation}
D(u^{\mu}u_{\mu}) / d\tau = 0
\end{equation}
which guarantees that [where we have chosen units where c=1 and
use the spacetime metric signature (1,-1,-1,-1)]
\begin{equation}
u^{\mu}u_{\mu} =1
\end{equation}
From this it follows that in general relativity the timelike
invariance of the world line of a massive particle is dynamically
preserved for all metric solutions, $g_{\mu\nu}$, to the Einstein
Equations, including the case of the "event horizon penetrating"
Kerr-Schild metric used by most black hole theoreticians in computer
simulations of the black hole collapse of a radially infalling
massive particle or fluid. On the basis of the above facts it is
then argued that there is no reason to look for physical
alternatives to black holes and that the assumption that the
preservation of the Strong Principle of Equivalence (SPOE) in Nature
implies that metrics with event horizons are non-physical, is in
error.

However we will now show that above arguments, which are based on
the four velocity $u^{\mu}$ alone, are not valid. This is because
relativists who come to this conclusion in this manner are making
the mistake of ignoring the fact that in addition to the four
velocity $u^{\mu}$ there exists another important quantity called
the "physical 3-velocity" which must also be considered as well.
Physically speaking, the magnitude of the physical 3-velocity is
seen by an observer at rest as being equal to the speed of the
co-moving observer who is moving along with the collapsing massive
particle or fluid. If we consider the case of a radially infalling
massive particle or fluid undergoing gravitational collapse it can
be shown that the radial component of the physical 3-velocity is
given by
\begin{equation}
V^r = c \frac{[(g_{0r}g_{0r} - g_{rr}g_{00})v^r
v^r]^{1/2}}{|(g_{00}+ g_{0r}v^r)|}
\end{equation}
where $v^r = dr/d\tau$ is the radial coordinate velocity of the
massive particle of fluid (See Landau and Lifshitz., 1975 "Classical
Theory of Fields", 4th Ed, Pergamon Press pg 248-252).

From the above formula for the radial component of the physical
3-velocity of the co-moving observer $V^r$ we see that for metrics
which have the property that $g_{00} \rightarrow 0$ in some region
of spacetime (i.e. the property associated with the existence of an
event horizon for the non-rotating metrics associated with the
radial infall of matter) the physical radial velocity $V^r$ of the
co-moving frame of the massive particle or fluid becomes equal to
the speed of light as the massive particle or fluid crosses the
event horizon.

Hence even though the timelike property $u^{\mu} u_{\mu} = 1$ is
preserved for a massive particle or fluid crossing the event horizon
of the Kerr-Schild metric where $g_{00} \rightarrow 0$ occurs, a
local special relativistic connection between the co-moving frame of
the radially infalling massive particle or fluid and a stationary
observer can no longer be made. Since the Strong Principle Of
Equivalence (SPOE) requires that Special Relativity must hold
locally at all points in spacetime, the breakdown at the event
horizon, of the local special relativistic connection between the
co-moving observer frame and a stationary observer frame for a
particle crossing the event horizon, represents a violation of the
SPOE. Hence we have shown that by considering both the four velocity and
the "physical 3-velocity of the co-moving observer" there is
motivation to look for physical alternatives to black holes. In fact in
this context logical arguments can be consistently made which show that
Black Holes with non-zero mass cannot exist in Nature
(Mitra, A., 2000, 2002, 2005, 2006).

Based on the above arguments, the SPOE
preserving requirement that the co-moving observer frame for a
massive collapsing fluid must always be able to be connected to a
stationary observer by special relativistic transformations with a
physical 3-speed which is less than the speed of light, was taken
seriously in our work. In the literature the requirement that the
SPOE must be preserved everywhere in spacetime for the timelike
worldlines of massive particles or fluids under the influence of
both gravitational and non-gravitational forces goes under the
technical name of "timelike worldline completeness".

Based on this idea we have found that preservation of the SPOE in
Nature can be accomplished only if there exist non-gravitational
components in the energy-momentum tensor on the right hand side of
the Einstein equation that physically guarantee the preservation of
the SPOE. It was in this alternative context that the general
relativistic MECO solutions to the Einstein-Maxwell equations
emerged, as was shown in the three previously published papers of
Robertson and Leiter and developed in more detail in Appendix 1-10
in this paper. There it was shown that for a collapsing body, the
structure and radiation transfer properties of the energy-momentum
tensor on the right hand side of the Einstein field equations, could
describe a collapsing radiating object which contained equipartition
magnetic fields that generated a highly redshifted Eddington limited
secular collapse process. This collapse process was shown to satisfy
the SPOE requirement of Timelike Worldline Completeness by
dynamically preventing trapped surfaces, that lead to event
horizons, from forming.

More specifically in Appendix 1-10 it was shown that, by using the
Einstein-Maxwell Equations and Quantum Electrodynamics in the
context of General Relativistic plasma astrophysics, it was possible
to virtually stop and maintain a slow, (many Hubble times!) steady
collapse of a compact physical plasma object outside of its
Schwarzschild radius. The non-gravitational force was Compton photon
pressure generated by synchrotron radiation from an intrinsic
equipartition magnetic dipole field contained within the compact
object. The rate of collapse is controlled by radiation at the local
Eddington limit, but from a highly redshifted surface. In Appendix 9
and 10 it was shown that general relativistic surface drift currents
within a pair plasma at the MECO surface can generate the required
magnetic fields. In Appendix 9 the equatorial poloidal magnetic
field associated with a locally Eddington limited secular rate of
collapse of the exterior surface was shown to be strong enough to
spontaneously create bound electron-positron pairs in the surface
plasma of the MECO. In the context of the MECO highly redshifted
Eddington limited balance, the action of this QED process was shown
to be sufficient to stabilize the collapse rate of the MECO surface.

For the case of hot collapsing radiating matter associated with the
MECO, the corresponding exterior solution to the Einstein equation was shown
to be described by the time dependent Vaidya metric. No
coordinate transformation between MECO Vaidya metric and the Black
Hole Kerr-Schild metric exists. Since the highly redshifted MECO
Vaidya metric solutions preserve the SPOE and they do not have event
horizons, they can also contain a slowly rotating intrinsic
magnetic dipole moment. These magnetic moments have observable effects if such
MECO exist at the centers of galactic black hole candidates and AGN.
In support of this idea our paper contains observational evidence
that the physical effects of such intrinsic magnetic dipole fields
in the central compact object in the Quasar Q0957 have been seen. It
is important to note that the physical effects of the MECO intrinsic
magnetic dipole fields that our observations have found in the
quasar Q0957 cannot be explained in terms of standard Black Hole
models using Kerr-Schild metric driven GRMHD calculations, since
these calculations generate unphysical "split magnetic monopole
fields" that cannot explain the details of the intrinsic structure
in Q0957 that our observations have found.

We reiterate that the Kerr-Schild metric
used by most relativists is not relevant to the work done in this
paper, because the collapsing radiating MECO solution to the
Einstein equation is described by the time dependent radiating
Vaidya metric, and there is no coordinate transformation between them.
Thus while we did not lightly ignore the Kerr-Schild metric
as applied to standard black hole solutions of the Einstein field
equations, our approach instead was to start with the gravitational
micro- and nano-lensing observations of the quasar Q0957, which
seemed to show the physical effects of an intrinsic dipole magnetic
field, attached to the collapsed slowly rotating central compact
object, which the Kerr-Schild metric solutions did not allow.
Therefore we found that we had to turn to alternate MECO Vaidya
metric solutions to the Einstein-Maxwell equations which feature the
intrinsic magnetic dipole fields implied by the observations.

\newpage
\begin{figure}
\begin{center}
\plotone{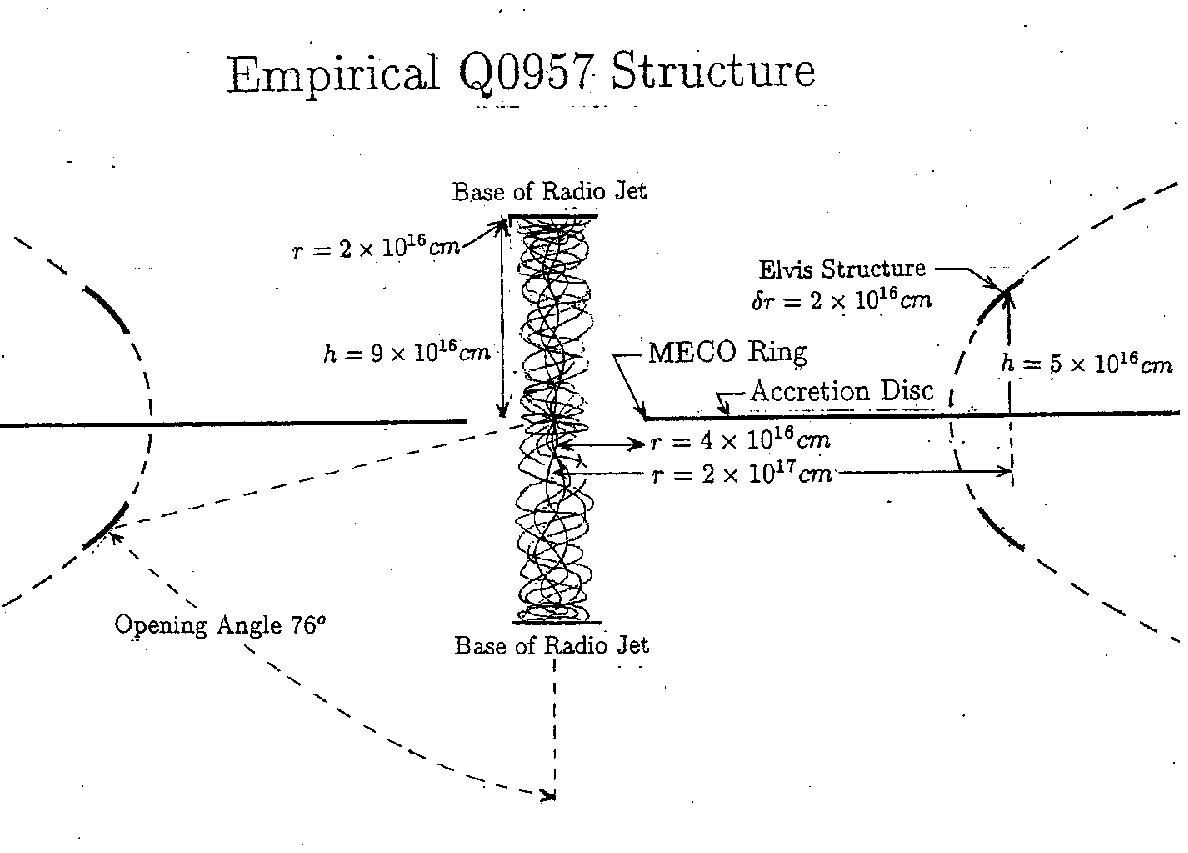}
\end{center}
\caption{A cartoon showing the principal structures in the empirical
quasar model and specifying the size scales of the structures found,
expressed in centimeters. In this cross-sectional view, the outer
Elvis outflow structures must be understood as a section of the
surfaces of revolution, and the luminous portions are shown as solid
line segments. These have been modeled as circles is the
Schild-Vakulik (2003) simulations. The inner luminous edge of the
accretion disc, marked MECO ring, has a very high surface
brightness, and is found to be outside an unexpectedly large empty
inner region. The tangled magnetic field lines are shown only to the
base of the radio jet. The overall geometry and proportions
correspond closely to the rotating magnetic star models of Romanova
et al (2002, 2003a).} \label{fig. 1}
\end{figure}
\newpage
\begin{figure} [t]
\begin{center}
\plotone{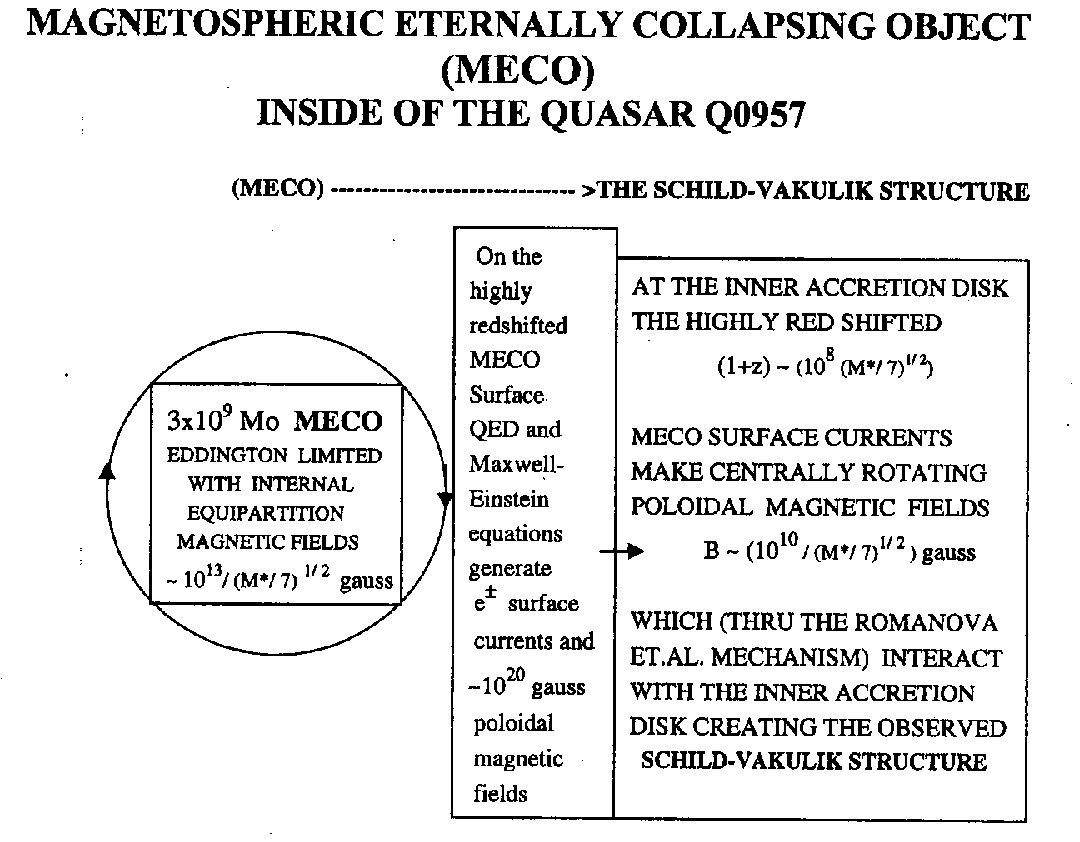}
\end{center}
\caption{ A logic diagram to show how the Magnetospheric Eternally
Collapsing Object (MECO) picture of physical quasar structure has been
applied to to the empirical quasar structure picture of Schild \& Vakulik
(2003). The rotating central object with Eddington limited internal
equipartition magnetic fields generates surface currents and
rotating poloidal
magnetic fields that clear out the inner quasar region. The central
rotating fields in the lo-hard state interact with the inner edge
of the accretion disc to produce the high surface brightness band at the
inner accretion disc edge. The magnetic fields operating through the
coronal region through the Romanova et al (2002, 2003a) interaction
create the outer Elvis outflow structures simulated in SV03. }
\label{fig. 2}
\end{figure}
 

\newpage
\begin{thebibliography}{}
\bibitem[]{2548} Abrams, L. S., 1979 Phys. Rev. D 20, 2474
\bibitem[]{2549} Abrams, L. S., 1989 Can J. Phys. 67, 919, gr-qc/0102051
\bibitem[]{2550}Arons, J. et al. 1984 in `High Energy Transients in
Astrophysics',
AIP Conf. Proc. 115, 215, Ed. S. Woosley, Santa Cruz, CA
\bibitem[]{2552}Baumgarte, T. \& Shapiro, S. 2003 ApJ 585, 930
\bibitem[]{2553}Bhattacharya D., \& Srinivasan, G., 1995 in `X-Ray
Binaries',
eds W. Lewin, J. van Paradijs \& E. van den Heuvel, Cambridge Univ.
Press
\bibitem[]{2556}Campana, S. et al., 2002 ApJ 580, 389
\bibitem[]{2557}Campana, S. et al., 1998 A\&A Rev. 8, 279
\bibitem[]{2558}Chou \& Tajima 1999 ApJ 513, 401
\bibitem{CS99} Colley, W. \& Schild, R., 1999, ApJ,
518, 153
\bibitem{CS00} Colley, W. \& Schild, R., 2000, ApJ,
540, 104
\bibitem{CS03} Colley, W. \& Schild, R.,
2003, ApJ, 594, 97 (CS03)
\bibitem{CS+} Colley, W. et al, 2003, ApJ,
587, 71
\bibitem[]{2567}Cui, W. 1997 ApJ 482, L163
\bibitem{DV05} De Vries et al, 2005, AJ, 129, 615
\bibitem{DMS} Diemand, J., Moore, B. \&
Stadel, J., 2005, Nature, 433, 389
\bibitem[]{2571}Done, C. \& $\dot{Z}$ycki, P. 1999 MNRAS 305, 457
\bibitem{ELV00} Elvis, M., 2000, ApJ, 545,
63
\bibitem{ELV94} Elvis, M. et al, 1994, ApJ
Supplements, 95, 1
\bibitem[]{2576}Gallo, E., Fender, R., Pooley, G., 2003 MNRAS, 344, 60
\bibitem[]{2577}Ghosh, P. \& Lamb, F. 1992 in X-Ray Binaries
and Recycled Pulsars, Ed. E. van den Huevel and S. Rappaport, Kluwer
\bibitem{G96} Gibson, C. 1996, Applied Mechanics review, 49 299;
astro-ph/9904260
\bibitem{G99} Gibson, C. 1999, astro-ph/9904260
\bibitem{GIV99} Giveon, U. et al, 1999, MNRAS, 306, 637
\bibitem[]{2583}Goodson, A., Bohm, K. \& Winglee, R. 1999 ApJ 534,
142
\bibitem{Gould97} Gould, A. and Miralde-Escude, J., 1997,
ApJ, 483, 13
\bibitem[]{2586}Gupta, A., Mishra, A., Mishra, H., \& Prasanna, A.R.,
1998,
`Classical
and Quantum Gravity', 15, 3131
\bibitem{H97} Haarsma, D. et al, 1997, ApJ, 479, 102
\bibitem{H99} Haarsma, D. et al, 1999, ApJ, 510, 64
\bibitem[]{2591}Harding, A., 2003, Invited talk at Pulsars, AXPs and
SGRs
Observed with BeppoSAX and Other Observatories, Marsala, Sicily,
Sept.
2002
astro-ph/0304120.
\bibitem{HAW96} Hawkins, M. 1996, MNRAS, 278, 787
\bibitem[]{2596}Hawley, J., Balbus, S. \& Winters, W., 1999 ApJ 518,
394
\bibitem[]{2597}Hernandez Jr., W.C. \& Misner, C.W.,
1966 ApJ. 143, 452
\bibitem{H04} Hill, A. et al, 2004, astro-ph/0411752
\bibitem{INA} Igumenschev, I. Narayan, R. and Abramowicz, M.,
2003, ApJ, 592, 1042
\bibitem[]{2602}Ilarianov, A. \& Sunyaev, R. 1975 A\&A 39, 185
\bibitem[]{2603}Kato, Y., Hayashi, M., Matsumoto, R. 2004 ApJ in press
astro-ph/0308437
\bibitem{K03} Koopmans, L. et al, 2003, ApJ, 595, 712
\bibitem{L92} Lehar, J. et al, 1992, ApJ, 384, 453
\bibitem{LR} Leiter, D. and Robertson, S. 2003, Found. Physics
Lett. 16, 143
\bibitem[]{2609}Lindquist, R. W., Schwartz, R. A. \&
Misner, C. W. 1965 Phys. Rev., 137B, 1364.
\bibitem[]{2611}Lindquist, R. W., 1966 Annals of Physics, 37, 487
\bibitem[]{2612}Maccarone, T., Gallo, E., Fender, R. 2003 MNRAS 345,
L19
\bibitem[]{2613}Markoff, S., Falcke H.,
\& Fender, R. 2001 A\&AL 372, 25
\bibitem[]{2615}Markoff, S., Nowak, M., Corbel, S., Fender, R., Falcke,
H.,
2003 New Astron. Rev. 47, 491
\bibitem[]{2617}Matt, S., Goodson, A., Winglee, R., B\"{o}hm, K. 2002
ApJ 574, 232
\bibitem[]{2618}Merloni, A., Fabian, A., 2002 MNRAS, 332, 165
\bibitem[]{2619}Merloni, A., Heinz, S., Di Matteo, T., 2003 MNRAS,
345, 1057
\bibitem[]{2620}Misner, C. W. 1965 Phys. Rev., 137B, 1360
2005-4-20rudyfile009.jpg\bibitem[]{2621}Misner, C., Thorne, K. \&
Wheeler, J. 1973 `Gravitation', Freeman,
San Francisco, California
\bibitem[]{2624}Mitra, A. 2000 Found. Phys. Lett. 13, 543
\bibitem[]{2625}Mitra, A., 2002 Found. Phys. Lett. 15, 439
\bibitem[]{2626}Mitra, A., 2005, Talk given at the 29th International Cosmic Ray Conference in
Pune India, Astro-ph/0506183
\bibitem[]{2627}Mitra, A., 2005 Astro-ph/0512006
\bibitem[]{2628}Mitra, A., in press 2006 in (Focus On Black Hole Research, ed. P.V. Kreitler,
Nova Science Publishers, Inc. ISBN 1-59454-460-3, novapublishers.com)
\bibitem{NQ} Narayan, R. and Quataert, E. 2005, Science, 307, 77
\bibitem{Osc01} Oscoz, A. et al 2001, ApJ, 552, 81
\bibitem[]{2628}Paragi, Z., et al. 2002 Talk presented
at the 4th Microquasar Workshop, Cargese, Corsica May 27-31 astro-
ph/0208125
\bibitem[]{2631}Pelletier, G. \& Marcowith, A., 1998 ApJ 502, 598
\bibitem{P96} Pelt, J. et al, 1996, A\&A, 305, 97
\bibitem{P98} Pelt, J. et al, 1998, A\&A, 336, 829
\bibitem{p97} Pijpers, F. 1997, MNRAS, 289, 933
\bibitem[]{2635}Possenti, A., Cerutti, R., Colpi, M.,
\& Mereghetti, S. 2002 A\&A 387, 993
\bibitem{R92} Press, W. et al, 1992, ApJ, 385, 404
\bibitem{PM02} Putman, M. and Moore, B. 2002, in
Extragalactic Gas at Low Redshift; ASP Conference
Proceedings 254, ed. J. Mulchaey and John Stocke
[San Francisco: Astronomical Society of the Pacific]
p. 245
\bibitem{RB91} Rauch, K. and Blandford, R., 1991,
ApJ, 381, 39
\bibitem{RS91} Refsdal, S., and Stabell, R.
1991, A\&A, 250, 62
\bibitem{RS93} Refsdal, S., and Stabell, R.
1993, A\&A, 278, L5
\bibitem{RS97} Refsdal, S., and Stabell, R.
1997, A\&A, 325, 877
\bibitem[]{2651}Reid, M. et al., 2003 ApJ 587, 208
\bibitem[gr]{GR04} Richards, G. et al, 2004, ApJ, 610, 679
\bibitem{RL02} Robertson, S., and Leiter, D.
2002, ApJ, 565, 447
\bibitem{RL03} Robertson, S., and Leiter, D.
2003, ApJ, 596, L203
\bibitem{RL04} Robertson, S., and Leiter, D.
2004, MNRAS, 350, 1391
\bibitem{RL05} Robertson, S., and Leiter, D.
2005, ``The Magnetospheric Eternally Collapsing Object (MECO)
Model of Galactic Black Hole Candidates and Active Galactic Nuclei'', pp
1-45 (in New Directions in Black Hole Research, ed. P.V.Kreitler,
Nova Science Publishers, Inc. ISBN 1-59454-460-3, novapublishers.com)
\bibitem{R02} Romanova, M. et al, 2002,
ApJ, 578, 420
\bibitem{R03a} Romanova, M. et al, 2003a,
ApJ, 588, 400
\bibitem{R03b} Romanova, M. et al, 2003b,
ApJ, 595, 1009
\bibitem{R04} Romanova, M. et al, 2004,
ApJ, 616, L151
\bibitem{S91} Schild, R., 1990, AJ, 100, 1991
\bibitem{S90} Schild, R. E. 1996, ApJ, 464,
125
\bibitem{S99} Schild, R. E. 1999, ApJ, 514,
598
\bibitem{S04b} Schild, R. E. 2004, astro-ph/0409549
\bibitem{S05} Schild, R. E. 2005, AJ, 129, 1225
\bibitem{SC86} Schild, R. E., \& Cholfin, B.
1986, ApJ, 300, 209
\bibitem{SS91} Schild, R. E., \& Smith,
R.C., 1991, AJ, 101, 813
\bibitem{ST97} Schild, R. E., \& Thomson, D.
J., 1995, AJ, 109, 1970
\bibitem{ST97} Schild, R. and Thomson, D.J. 1997, AJ,
113, 130
\bibitem{ST97a} Schild, R. E., \& Thomson,
D. J., 1997a, AJ, 113, 130
\bibitem{ST97b} Schild, R. E., \& Thomson,
D. J. 1997b, in Astronomical
Time Series, ed. D. Maoz, A.
Sternberg, and E. Liebowitz
[Boston: Kluwer] p. 74
D. J., 1995, in Dark Matter: AIP
S. Holt \& C. Bennett [AIP Press:
\bibitem{SV03} Schild, R. E., \& Vakulik,
V., 2003, AJ, 126, 689 (SV03)
\bibitem{TS97} Thomson, D. J., \& Schild,
R., 1997,, in Applications of
Time Series Analysis in Astronomy
and Meteorology, ed. T. Subba
Rao, M. Priestly, \& O. Lessi
[Chapman and Hall: New York], 187
\bibitem[]{2701}Shapiro, S. \& Teukolsky, S. 1983 in
`Black Holes, White Dwarfs \& Neutron Stars', John Wiley \& Sons, Inc.,
New York
\bibitem{SGH} Spekkens, K, Giovanelli, R. and Haynes, M. 2005, AJ,
129, 2119
\bibitem[]{2705}Stella, L., White, N. \& Rosner, R. 1986 ApJ 308, 669
\bibitem[]{2706}Thorne, K., 1965 Phys. Rev, 138, B251
\bibitem[]{2707}Uzdensky, D., 2002 ApJ 572, 432
\bibitem[]{VA06}Vakulik, V., 2006, A\&A, in press; astro-ph/0509545
\bibitem{WW98} Walker, M. and Wardle, M., 1998,
ApJ, 498, 125
\bibitem{WW99b} Wardle, M. and Walker, M., 1999,
ApJ, 527, 109
\bibitem[]{2712}Wheeler, J. \& Ciufolini, I., 1995 in `Gravitation And
Inertia'
Princeton Univ. Press, 41 William St., Princeton, New Jersey
\bibitem[]{2714}White, N.\& Marshall, F. 1984, ApJ. 281, 354
\bibitem[]{2715}Wilson, C. \& Done, C. 2001 MNRAS 325, 167
\bibitem{W04} Winn, J. et al, 2004, AJ, 128, 2696
\bibitem[WY]{Wy0} Wyithe, S. webster, R. and Turner, E., 2000, MNRAS, 318,
1120
\bibitem[]{2717}Zaumen, W. T. , 1976 ApJ, 210, 776
\bibitem[]{2718}Zhang, W., Yu, W. \& Zhang, S. 1998 ApJ 494, L71
\bibitem[]{2719}$\dot{Z}$ycki, P., Done, C. and Smith, D. 1997a in AIP
Conf. Proc. 431, Accretion Processes in Astrophysical Systems:
Some Like It Hot, Ed. S. S. Holt \& T. R. Kallman (New York, AIP),
319
\bibitem[]{2722}$\dot{Z}$ycki, P., Done, C. and Smith 1997b ApJ 488,
L113
\end{thebibliography}
\end{document}